\documentclass[useAMS,usenatbib,usegraphicx]{mn2e}

\newcommand{\apj}{ApJ} 
\newcommand{\apjl}{ApJ} 
\newcommand{\aap}{A\&A} 
\newcommand{\aj}{AJ} 
\newcommand{\apjs}{ApJS} 
\newcommand{\mnras}{MNRAS} 
\newcommand{\pasp}{PASP} 
\renewcommand{\d}{\ensuremath{\mathrm{d}}}

\title[K-band galaxies from UKIDSS LAS]{Luminosity and surface brightness distribution of $K$-band galaxies from the UKIDSS Large Area Survey}
\author[A.\ J.\ Smith,  J.\ Loveday \& N.\ J.\ G.\ Cross]
	{Anthony J. Smith$^{1}$\thanks{E-mail: \href{mailto:A.J.Smith@sussex.ac.uk}{\texttt{A.J.Smith@sussex.ac.uk}}},
	Jon  Loveday$^{1}$ and Nicholas J. G. Cross$^{2}$\\ 
	$^{1}$Astronomy Centre, University of Sussex, Falmer, Brighton BN1 9QH\\
	$^{2}$Scottish Universities Physics Alliance, Institute for Astronomy, University of Edinburgh, Royal Observatory, Edinburgh EH9 3HJ}

\usepackage{amsmath}
\usepackage{amssymb}
\usepackage[colorlinks,pdfusetitle,citecolor=black,urlcolor=black,linkcolor=black]{hyperref}

\begin{document}

\date{Accepted 2009 April 27. Received 2009 April 21; in original form 2008 June 2}

\pagerange{\pageref{firstpage}--\pageref{lastpage}} \pubyear{2009}

\maketitle

\label{firstpage}


\begin{abstract}
We present luminosity and surface-brightness distributions of 40\,111 galaxies with $K$-band photometry from the United Kingdom Infrared Telescope (UKIRT) Infrared Deep Sky Survey (UKIDSS) Large Area Survey (LAS), Data Release 3 and optical photometry from Data Release 5 of the Sloan Digital Sky Survey (SDSS). Various features and limitations of the new UKIDSS data are examined, such as a problem affecting Petrosian magnitudes of extended sources. Selection limits in $K$- and $r$-band magnitude, $K$-band surface brightness and $K$-band radius are included explicitly in the $1/V_\mathrm{max}$ estimate of the space density and luminosity function. The bivariate brightness distribution in $K$-band absolute magnitude and surface brightness is presented and found to display a clear luminosity--surface brightness correlation that flattens at high luminosity and broadens at low luminosity, consistent with similar analyses at optical wavelengths. Best fitting Schechter function parameters for the $K$-band luminosity function are found to be $M^*-5\log h=-23.19 \pm 0.04$, $\alpha=-0.81 \pm 0.04$ and $\phi^*=(0.0166 \pm 0.0008)h^3$\,Mpc$^{-3}$, although the Schechter function provides a poor fit to the data at high and low luminosity, while the luminosity density in the $K$ band is found to be $j = (6.305 \pm 0.067) \times 10^8$\,L$_\odot\,h$\,Mpc$^{-3}$. However, we caution that there are various known sources of incompleteness and uncertainty in our results. Using mass-to-light ratios determined from the optical colours we estimate the stellar mass function, finding good agreement with previous results. Possible improvements are discussed that could be implemented when extending this analysis to the full LAS.
\end{abstract}


\section{Introduction}

It is possible to learn much about a population's history by taking a census of the present-day population. With a large and increasing number of deep, large-area surveys taking place, a census of the low-redshift galaxy population may be undertaken. Deep imaging allows many different properties to be studied simultaneously.  Much can be learned about galaxy formation and evolution by investigating these properties: how they correlate with each other, and the sub-populations that exist.

The advantages of working at low redshift ($z \sim 0.1$) are as follows: (1) a more complete sample may be studied, including galaxies with low luminosity or low surface brightness, (2) the more luminous galaxies may be studied in more depth, investigating morphology and structure as well as luminosity and colour and (3) the evolution and selection effects that plague high-redshift surveys are less of a problem.

The advantages (in principle) of building such a census on near-infrared (NIR) observations are well known. First, mass-to-light ($M/L$) ratios in the NIR are largely insensitive to galaxy or stellar type, certainly much less than in the optical \citep{BelldJ2001}. This means that the NIR light is a good tracer of the total stellar mass in a galaxy. Moreover, the range of $M/L$ ratios is much smaller in the NIR, so uncertainties in the stellar mass are much smaller. Not only does this mean that a survey limited in NIR magnitude will be approximately limited in `apparent stellar mass', but also that morphological measurements in the NIR, for example the S\'ersic index and the half-light radius, will reflect the distribution of stellar mass within the galaxy, whereas such measures in the optical will be significantly biased by the presence of young stellar populations. The $K$-band galaxy luminosity function (LF) is, for these reasons, a convenient quantity for numerical or semi-analytic models to predict: see e.g.\ \citet{Croton...2006}, \citet{Bower...2006}, \citet{deLuciaB2007} and \citet*{BertonedLT2007}.

A second advantage is that the $K$-corrections in the $K$ band are also relatively independent of galaxy type \mbox{\citep{Mannucci...2001}}, leading to smaller uncertainties in the absolute magnitudes.

A third advantage is that dust is much less of a problem in the NIR than in the optical.  This means that, whereas optical measurements of galaxy properties are affected by dust obscuration, and therefore strongly dependent on the inclination of the galaxy, producing a smoothing of the galaxy LF, this is not such a problem in the NIR \citep{Driver...2007b,Maller...2009}.

However, the main disadvantage of the NIR (for ground-based telescopes) is the sky brightness, which is around 13.5\,mag\,arcsec$^{-2}$ in $K$ for the data used here \citep{Dye...2006}.

There have been several studies of the low-redshift $K$-band LF using Two-Micron All-Sky Survey (2MASS) imaging combined with various redshift surveys \citep{Cole...2001,Kochanek...2001b,Bell...2003c,Eke...2005,Jones...2006}. Other estimates have been made by \citet*{MobasherSE1993}, \citet{Szokoly...1998} and \citet{Loveday2000}, using optically selected samples, and by \citet{Glazebrook...1995b}, \citet{Gardner...1997} and \citet{Huang...2003}, using samples selected in the $K$ band. Table \ref{tbl:samplesize} shows the sample size of these $K$-band LF estimates.

An accurate detemination of the local $K$-band LF also has great value as a baseline for comparison with studies at higher redshift \citep[e.g.][]{Cirasuolo...2007}. The principal uncertainty remaining at low redshift is connected with the low-luminosity end of the LF.  Moreover, there has been significant discussion about possible low-surface-brightness incompleteness in 2MASS \citep{Andreon2002a}, which would affect the low-luminosity end of the LF.

\begin{table}
\caption[Sample sizes of $K$-band galaxy LFs]{\label{tbl:samplesize} Sample sizes of $K$-band galaxy luminosity functions.}
\vspace{.2in}
\centering
\begin{tabular}{lr}
\hline
\hline
Paper & Number of galaxies in sample \\
\hline
\citet{MobasherSE1993} & 181 \\
\citet{Glazebrook...1995b} & 124 \\
\citet{Gardner...1997} & 510 \\
\citet{Szokoly...1998} & 175 \\
\citet{Loveday2000} & 345 \\
\citet{Kochanek...2001b} & 3878 \\
\citet{Cole...2001} & 5683 \\
\citet{Huang...2003} & 1056 \\
\citet{Bell...2003c} & 6282 \\
\citet{Eke...2005} & 15\,644 \\
\citet{Jones...2006} & 60\,869 \\
This work & 40\,111 \\
\hline
\end{tabular}
\end{table}

Here we present the first statistical study of galaxies in the UKIRT Infrared Deep Sky Survey (UKIDSS) Large Area Survey, while leaving detailed investigation of surface brightness completeness and the very faint-end of the LF to future work.

This paper is organized as follows. Details about the sample used are found in Section 2. The method of analysing the data is described in Section 3. In Section 4, the bivariate brightness distribution (BBD) and LF are presented. The stellar mass function (SMF) is estimated in Section 5. There is a discussion in Section 6, followed by the conclusions in Section 7.

For ease of comparison with previous results, a flat cosmological model with $\Omega_\mathrm{M}=0.3$ and $\Omega_\Lambda=0.7$ is used, with $H_0 = 100h$\,km\,s$^{-1}$\,Mpc$^{-1}$. AB magnitudes are used for Sloan Digital Sky Survey (SDSS) magnitudes and Vega magnitudes for $K$-band quantities.  For reference, AB and Vega magnitudes are related in the $r$ band by $r_\mathrm{AB} = r_\mathrm{Vega} + 0.146$ and in the $K$ band by $K_\mathrm{AB} = K_\mathrm{Vega} + 1.900$ \citep{Hewett...2006}.

\section{Data}

All of the galaxies used in this sample are drawn from the main galaxy sample of Data Release 5 (DR5) of the SDSS \citep{Adelman-McCarthy...2007}, from which the optical photometry and spectroscopic redshifts used in the analysis below are obtained.  The imaging survey covers 8000 square degrees, yielding a sample of 783\,070 target galaxies. As of DR5, the spectroscopic sample had covered 5740 deg$^2$ of the imaging area, with high redshift completeness.

The UKIRT Infrared Deep Sky Survey (UKIDSS) is defined by \citet{Lawrence...2007}. UKIDSS uses the UKIRT Wide Field Camera \citep[WFCAM;][]{Casali...2007} and a photometric system described by \citet{Hewett...2006}. The pipeline processing and science archive are described by Irwin et al.\ (in preparation) and \citet{Hambly...2008}, respectively.

UKIDSS consists of five surveys at a variety of depths and areas and using various combinations of the $ZYJHK$ filters.  $K$-band data for this paper are taken from the UKIDSS LAS, which is contained within the field of SDSS.  UKIDSS Data Release 3 (DR3; Warren et al., in preparation), which we have used here, was released in 2007 December, with the LAS containing coverage in $YJHK$, including 1189 deg$^2$ of coverage in $K$ to a 5$\sigma$ depth of 18.2\,mag, in directions towards both the north and south Galactic poles (NGP and SGP, respectively).

\subsection{Area covered by the UKIDSS--SDSS sample}
\label{sec:area}

The volume, and hence the area, must be well estimated for the normalization of the LF and related quantities. Given the complex geometry of the overlap region between the two surveys, we estimate the area by dividing the number of galaxies in the matched sample by the number density of sources (per square degree) in the SDSS sample.

A substantial fraction of the region of overlap between the two samples was surveyed in the SDSS Early Data Release, so we use the limit of $r < 17.6$ (de-reddened Petrosian magnitude) rather than $r < 17.77$ used in later versions of the SDSS main galaxy selection algorithm. This corresponds to 635\,320 target galaxies over the whole area.

The number density is estimated using the total area of the SDSS imaging survey (8000 deg$^2$) and the number of galaxies targeted for spectroscopy in the SDSS main galaxy sample. But, first the size of this sample must be corrected for those objects included that are not galaxies. In SDSS DR5, the number of target galaxies with good spectroscopic redshifts (zConf $>0.8$) and with de-reddened Petrosian magnitudes brighter than $r = 17.6$ is 391\,052. Of these, 384\,617 are spectroscopically classified as galaxies. This suggests that around $(391\,052-384\,617)/391\,052 = 1.6$ per cent of the target galaxies are not in fact galaxies. Taking this into account gives a corrected target sample size of 624\,865, giving a source density of 78.11 galaxies per square degree.

\begin{figure}
\centerline {
\includegraphics[width=0.5\textwidth]{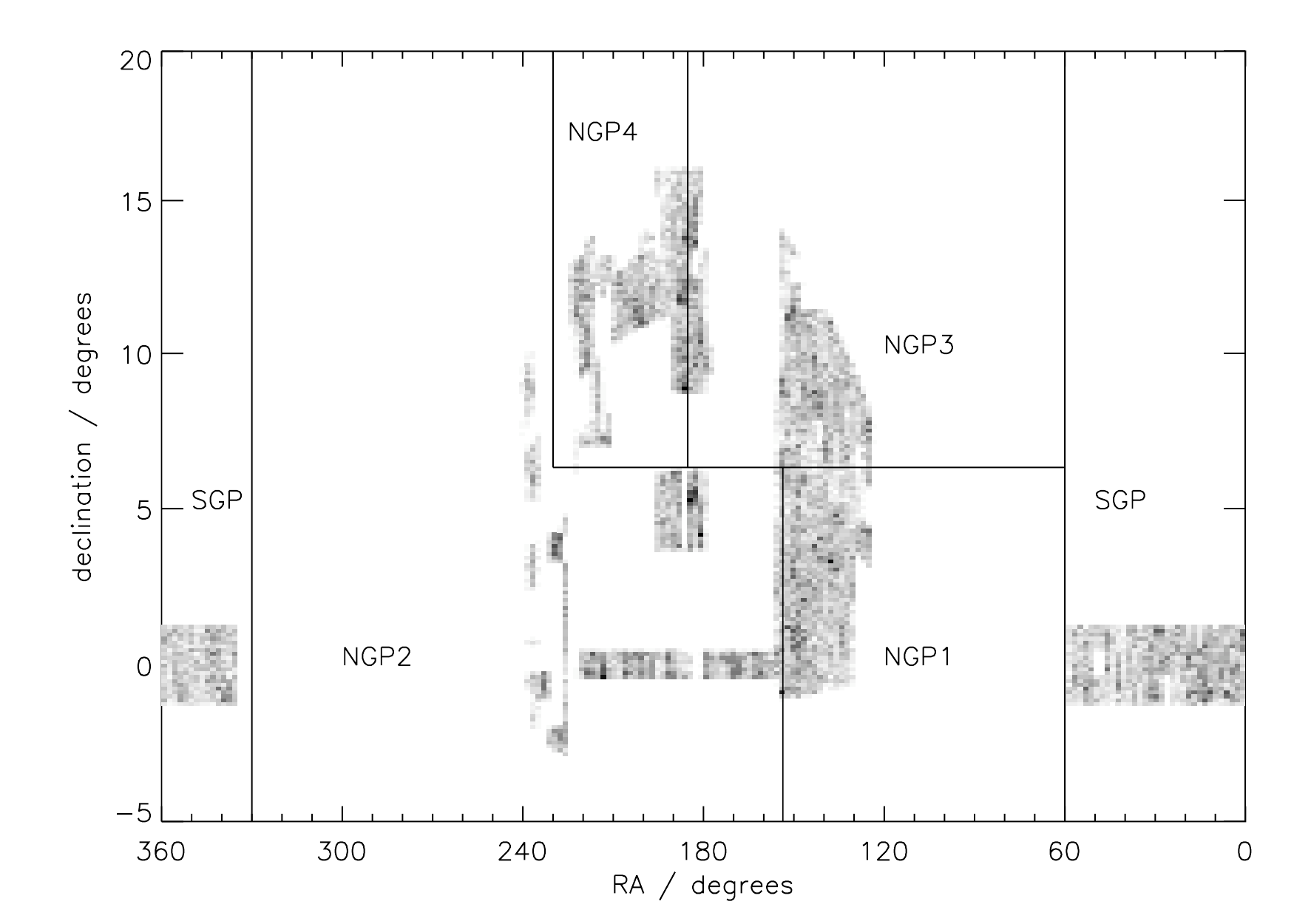}}
\caption[coverage]{\label{coverage} Sky coverage of the sample, showing the principal regions used for the jackknife samples (see Section \ref{jackknife}), each of which is further subdivided into strips in RA (four in the NGP4 region and five in the others) giving 24 jackknife regions in total, each containing approximately the same number of galaxies. In the regions with coverage, a darker shade of grey corresponds to a higher number density of galaxies.}
\end{figure}

The two data sets were matched using the WFCAM Science Archive (WSA).\footnote{\url{http://surveys.roe.ac.uk/wsa/}} The initial sample was found by selecting all closest matches within 2 arcsec with good, non-zero spectroscopic redshifts (SDSS zConf $>0.8$), classified spectroscopically as galaxies and having no major quality control issues flagged in $K$ (UKIDSS kppErrBits $<256$), yielding a sample of 49\,255 galaxies. The sky coverage of this matched sample is shown in Fig.\ \ref{coverage}. 

This number will be affected by failed detections, which may introduce some bias into the sample. One type of failed detection is when redshifts have not been obtained, due to lack of coverage, failed redshifts or lack of available fibres to measure the spectra (`fibre collisions'). These are assumed to introduce no bias into the sample, although it has been noted \citep{Blanton...2003a, Blanton...2005c} that the SDSS fiber collisions lead to slight incompleteness at high-density regions, which may introduce a small bias against the type of galaxy found in such environments.

\begin{figure}
\centerline {
\includegraphics[width=0.5\textwidth]{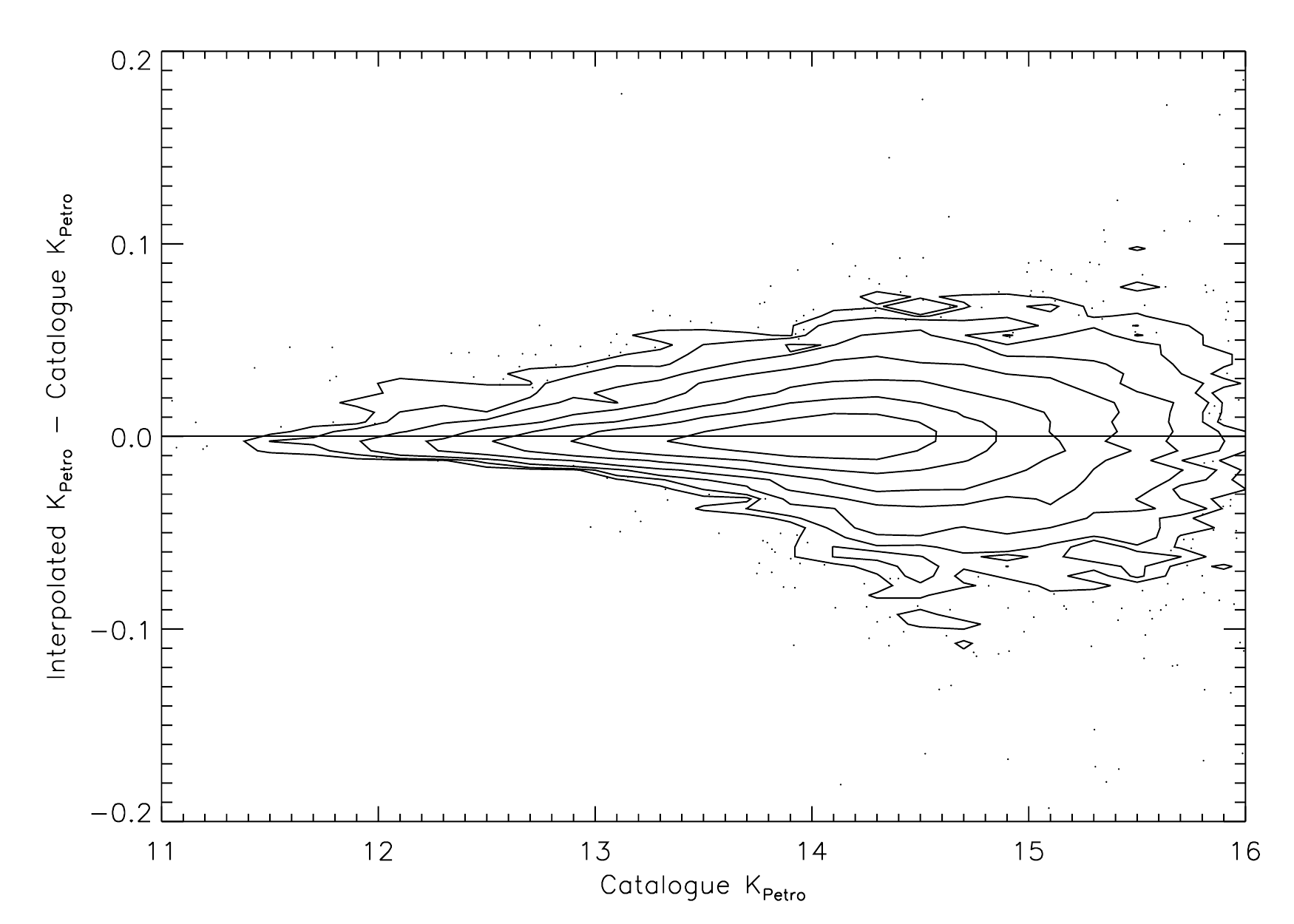}}
\caption[kpetro_corrected]{\label{kpetro_corrected} A comparison between the catalogue $K$-band Petrosian magnitudes and the magnitudes estimated by interpolating between the fixed aperture fluxes, for the 36\,660 galaxies in the final sample that are not flagged as deblended. Contours show the density of sources on a logarithmic scale, with sources shown as points where the density is low. The mean offset is set to be zero and the standard deviation is 0.018\,mag.}
\end{figure}

Another kind of failed detection is when there is a problem with the UKIDSS catalogue extraction. During the course of this work, a bug in the source extraction was discovered affecting the deblending algorithm. It was found that, for Petrosian magnitudes in $Y$, $H$ and $K$, the deblender, when invoked, was making the source significantly brighter. (The $J$-band data are micro-stepped, unlike $Y$, $H$ and $K$, which may explain why this problem is not seen for $J$-band Petrosian magnitudes.) This bug is in the process of being fixed; the reader is referred to the WSA web site for updates.\footnote{\url{http://surveys.roe.ac.uk/wsa/knownIssues.html}} However, in the present analysis, for the 4835 sources flagged as deblended (almost 10 per cent of the matched sample), we have estimated the flux within the Petrosian aperture by quadratic interpolation between the fixed-aperture fluxes, taking advantage of the fact that the Petrosian \textit{radii}, $r_\mathrm{P}$, have not been seriously affected by the bug \citep[note that `even substantial errors in $r_\mathrm{P}$ cause only small errors in the Petrosian flux';][]{Blanton...2001}.  By comparing the catalogue $K$-band Petrosian magnitudes for non-deblended sources with those estimated by this method, we find a scatter of 0.018\,mag, which suggests that the method is sufficiently accurate to provide Petrosian magnitudes for those sources affected by deblending. This is shown in Fig.\ \ref{kpetro_corrected}.

Of the 49\,255 sources in the matched sample, 46 sources are flagged as having bad pixel(s) in the default aperture. These have been removed, leaving 49\,209 in the sample.

\begin{figure}
\centerline {
\includegraphics[width=0.5\textwidth]{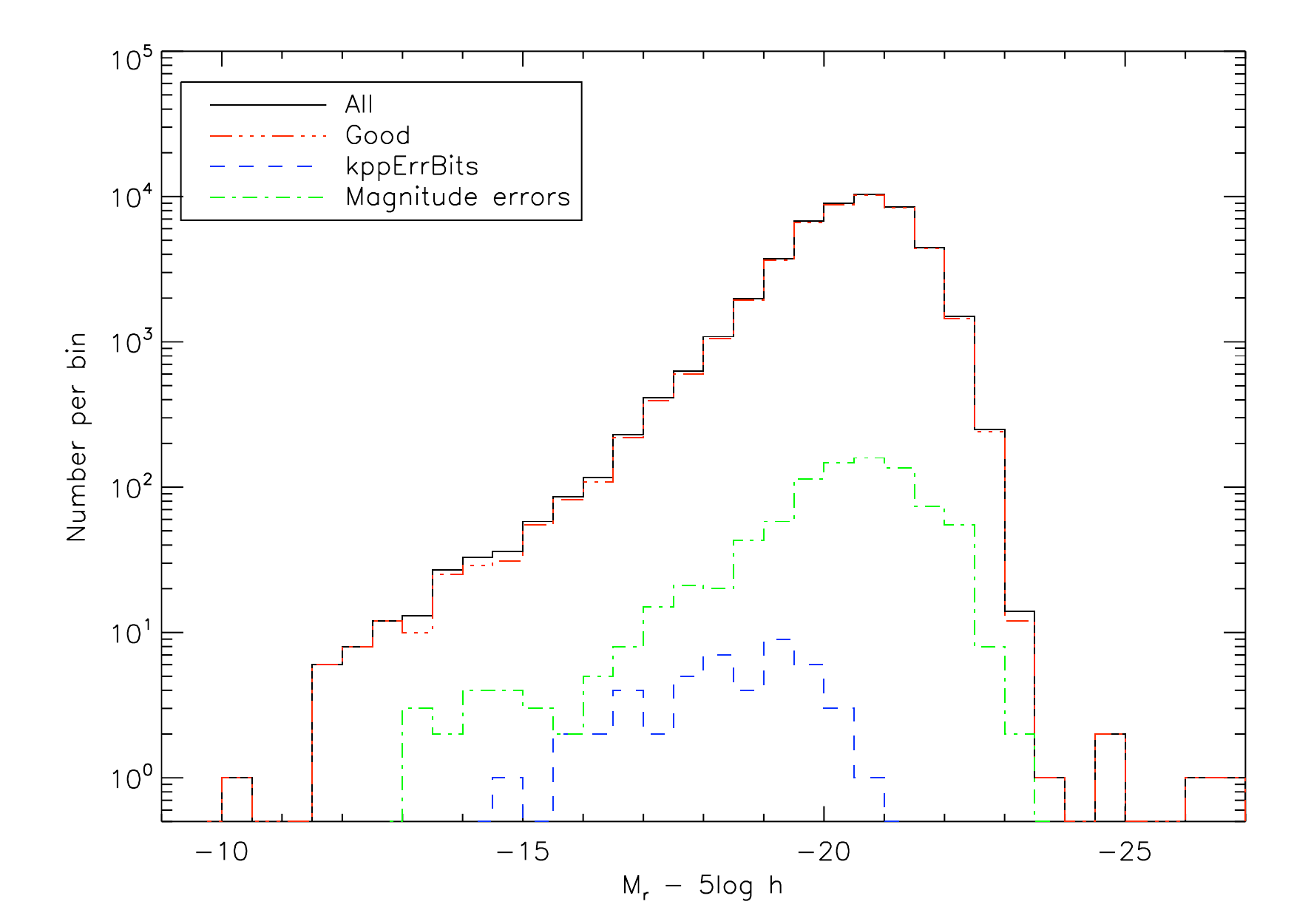}}
\caption[cut_bias]{\label{cut_bias} $r$-band absolute magnitude of the sources in the entire sample (`All', 49\,255), the final matched sample (`Good', 48\,327), those excluded due to poor $K$-band imaging, i.e.\ bad pixels (`kppErrBits', 46) and those excluded due to uncertainties in the Petrosian magnitudes (`Magnitude errors', 882).}
\end{figure}

Fig.\ \ref{cut_bias} shows the $r$-band absolute magnitude of galaxies in the whole sample, and of sources excluded because of problems with the imaging.

A small number of the remaining galaxies have very large magnitude errors, greater or much greater than 0.15\,mag. So, one final cut is to remove sources with large uncertainty in magnitude, in order to restrict the systematic errors in our results. Of the 49\,209 galaxies remaining, those with magnitude errors greater than 0.15\,mag in $r$ or $K$ are removed (69 in $K$, 811 in $r$ and 2 in both) leaving 48\,327 galaxies. Given the small number affected by this cut, and from Fig.\ \ref{cut_bias}, it is assumed that any bias induced by this cut will be negligible.

By estimating the area in this way, the assumption is that all SDSS target galaxies would be detected in the LAS, if that part of the sky has been surveyed.  If this is not the case, it will have two effects: (1) particular types of galaxies will be underrepresented in the sample (those within the SDSS completeness limits but outside the limits for the LAS) and (2) the overall normalization will be too high, as the area and hence the volume probed will be underestimated.

\begin{figure}
\centerline {
\includegraphics[width=0.5\textwidth]{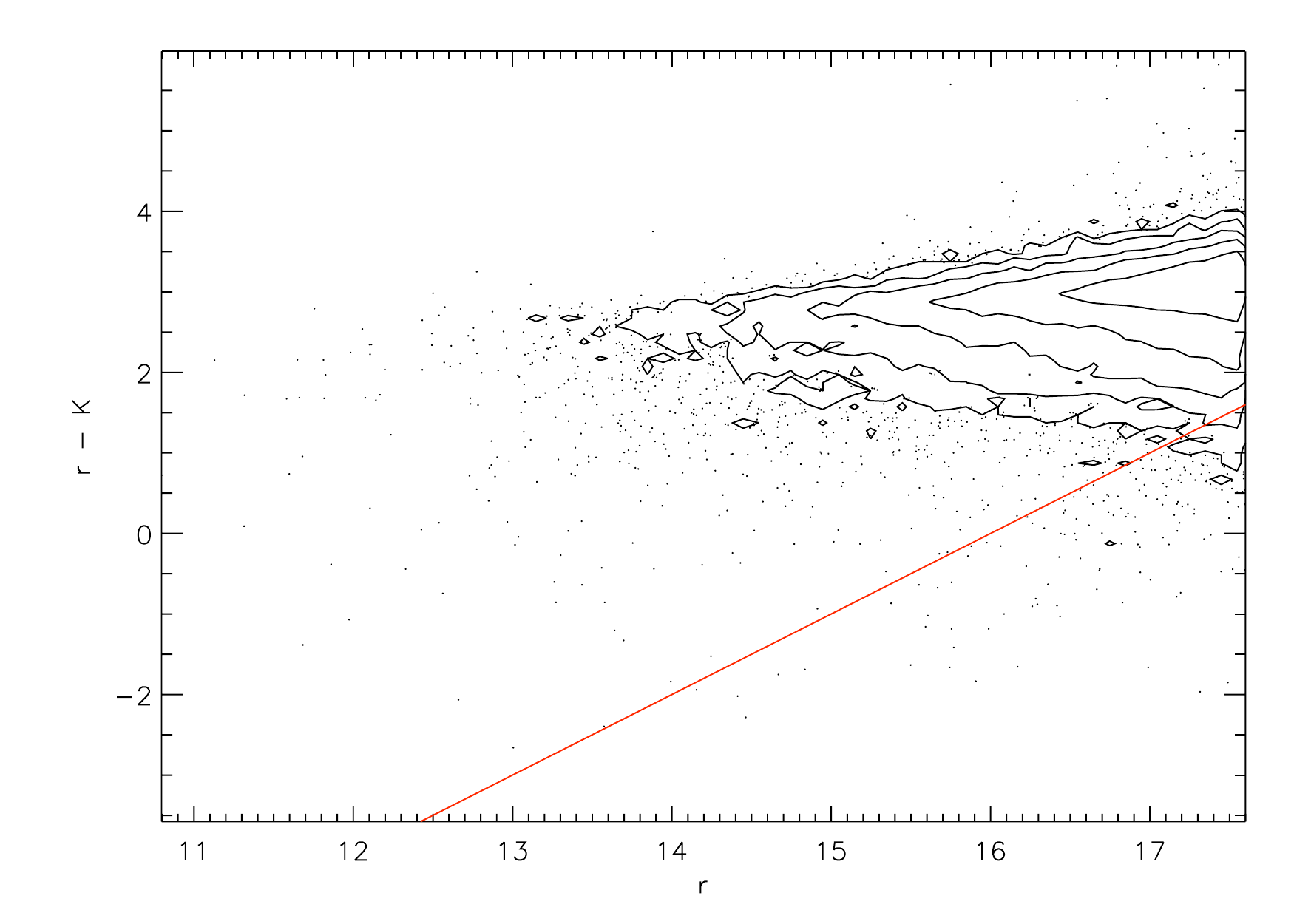}}
\caption[rmk]{\label{rmk} $r$-band Petrosian magnitude and $r-K$ Petrosian colour for 48\,327 galaxies in the matched sample, before imposing a cut at $K < 16$. Objects lying below the solid (red) line have $K > 16$ and will not be included in the final analysis. Contours are spaced on a logarithmic scale, with sources shown as points where the density is low.}
\end{figure}

Fig.\ \ref{rmk} shows $r$-band Petrosian magnitude and the $r-K$ Petrosian colour (note that the apertures are not the same in $r$ and $K$ so this is not a true colour) for the sources in the matched sample. From the figure it can be seen that (1) there are likely to be very few sources at all lying within the SDSS flux limit but outside the $K$-band limit, and (2) many of these sources are detected anyway, since there are many sources detected fainter than the (nominal) $K$-band limit.  This suggests that the effect on the overall normalization will be negligible, although the colour-dependent bias will be considered later.

The effective area can now be given as $48\,327 / 78.11 = 618.72$\,deg$^2$.

Calculating the area in this way takes into account any other \textit{random} (not bias-inducing) sources of incompleteness that have not been considered here explicitly.

Note that the final sample will be smaller yet, due to limits imposed on magnitudes, radius and surface brightness, but this is related to the redshift limits (which affect the volume probed) rather than the area covered.

\subsection{Apparent magnitudes}

\citet{Petrosian1976} apparent magnitudes are used here \citep[for comparison with isophotal magnitudes, see][]{Blanton...2001} and Galactic extinction corrections are used throughout, based on the extinction maps of \citet*{SchlegelFD1998}.

\begin{figure}
\centerline {
\includegraphics[width=0.5\textwidth]{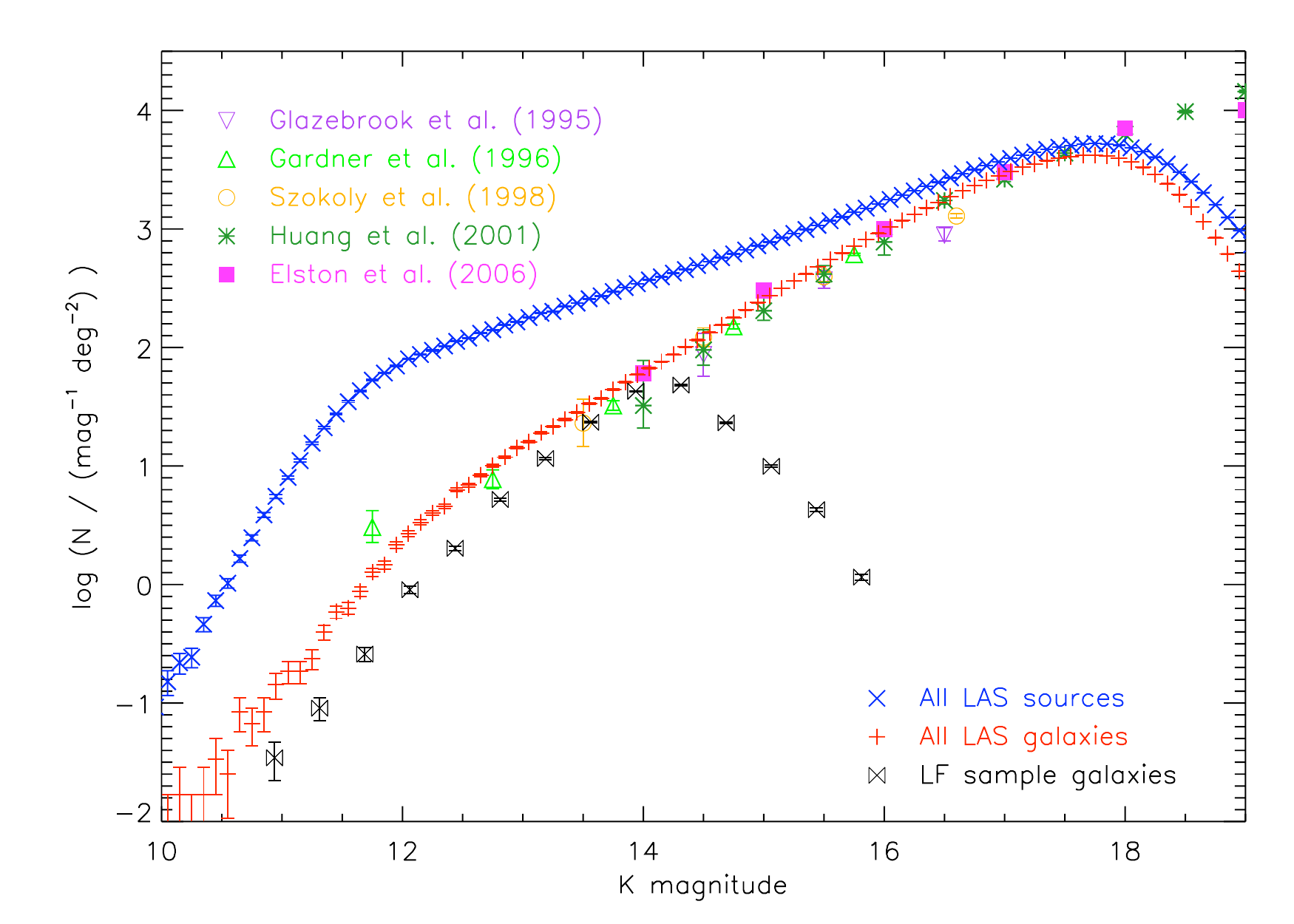}}
\caption[k_counts]{\label{k_counts} $K$-band number counts for the full DR3 LAS sample (requiring a seamless sample, with no quality control error bits flagged), for the final LF sample, and for various samples from the literature. The blue crosses show the number counts for the whole sample (15\,079\,199 objects), while the red plus-signs are for sources classified as galaxies (kClass $=1$; 9\,962\,258 objects). Poisson error bars are shown. It may be seen that the counts for the LF sample lie always below the counts for the whole UKIDSS sample and have a different slope, suggesting that, at all magnitudes, the sample is limited by other factors in addition to the $K$-band magnitude. \nocite{Glazebrook...1995b,Gardner...1996,Szokoly...1998,Huang...2001,Elston...2006}}
\end{figure}

Fig.\ \ref{k_counts} shows the number counts as a function of $K$-band Petrosian magnitude for the whole UKIDSS LAS DR3. Comparison with previous results shows evidence of incompleteness in the galaxy number counts fainter than $K=17$. We choose to limit the sample to $K < 16$. We use a faint magnitude limit of 17.6 in $r$.  Corrections are not applied to compensate for the effect of seeing on the Petrosian magnitudes, or for the different fraction of the galaxy's flux recovered by the Petrosian magnitudes for different galaxy types \citep[99 per cent for an exponential profile and 82 per cent, or $+0.22$\,mag, for a de Vaucouleurs profile;][]{Blanton...2001}.

Bright limits are applied to the SDSS fibre magnitudes, since the spectroscopic sample requires these to be fainter than 15 in $g$ and fainter than 14.5 in $i$.  These, however, have no effect on our results since the strongest constraint on the low-redshift visibility of each galaxy is generally the large radius limit.  No bright limit has been applied in the $K$-band; adding a limit as strong as $K>13$ has a barely noticeable effect.

\subsection{Radii}
\label{sec:radii}

Two radii are used below: the Petrosian radius and, for calculating the surface brightness, the elliptical half-light (semi-major axis) radius.

Given the sky brightness in the $K$-band and the depth of the LAS, the sky subtraction algorithm has to operate with a relatively small pixel size of around 24 arcsec.  This is set in the UKIDSS pipeline as the maximum diameter for the (circular) aperture over which the galaxy's flux is integrated.  A diameter of 24 arcsec corresponds to a Petrosian radius of 6 arcsec, so this means that any source with a true Petrosian radius greater than 6 arcsec will have its radius clipped at this value, and the quantity labelled as the Petrosian magnitude will in fact be an underestimate of the Petrosian flux.

To avoid complication, a large radius limit has been set, requiring the Petrosian radius to be less than 6 arcsec. It is worth noting that a significant fraction of the sample (7513 out of 48\,327) have their Petrosian radii clipped at 6 arcsec. This limit is taken into account when estimating the contribution of each galaxy to the space density. However, when the large-radius limit is combined with the maximum redshift limit of $z=0.3$ (see Section \ref{sec:redshift}), it will have the effect of excluding all galaxies above a certain physical size. With the angular diameter distance, $D_\mathrm{A}(z)$, given in $h^{-1}$\,Mpc, the Petrosian radius of the largest galaxy that could be visible in the sample is
\begin{eqnarray}
\label{eqn:radius}
R = \frac{1000\pi (6\,\mbox{arcsec}) D_\mathrm{A}(0.3)}{ 180 \times 3600} = 18.7h^{-1}\,\mbox{kpc} \,.
\end{eqnarray}
This will lead to an incompleteness of very large galaxies in the final results.

No small radius limit has been set, as the problem of misclassifying galaxies as stars is thought to be negligible in SDSS \citep{Blanton...2001}, and we do not require independent classification in the $K$ band.

The scale-size of a galaxy is conventionally measured using the radius enclosing half of the total light, known as the half-light or effective radius and denoted by $r_\mathrm{e}$. In order to correct for inclination, the half-light radius we would like is the semi-major axis of the elliptical aperture, of the same ellipticity and position angle as the galaxy, containing half the total flux of the galaxy. This is estimated (Cross et al., in preparation) using the Petrosian flux, the 13 circular aperture fluxes, the ellipticity and the seeing, all made available in the WSA (the pipeline does not measure the half-light radius).

\subsection{Surface brightness}

The half-light, or effective, surface brightness is estimated from the half-light radius by \citep{Blanton...2001}
\begin{eqnarray}
\label{sb}
\mu_\mathrm{e} = m + 2.5 \log 2 \pi r_\mathrm{e}^2
\end{eqnarray}
where $m$ is the Petrosian magnitude.

Tests on the UKIDSS source extraction (Cross et al., in preparation) suggest that, for a de Vaucouleurs profile, galaxies with surface brightness fainter than 19.5\,mag\,arcsec$^{-2}$ are likely to have their fluxes and sizes underestimated. For well-defined sample limits, we should impose a cut in surface brightness at that value. However, when investigating the space density of galaxies with high surface brightness, this limit can be safely ignored, since for the vast majority of the sample, the faint magnitude limit in $r$ provides a stronger constraint on the visibility of the galaxy.  So in order to include at least some low-surface brightness galaxies in the analysis we set a limit in the $K$-band of $\mu_\mathrm{e} < 21$\,mag\,arcsec$^{-2}$.

The combined limits in $K$-band Petrosian magnitude and Petrosian radius will impose a limit on the effective surface brightness for each galaxy.  For \citet{Sersic1968} indices $n$ between 1 and 4, the Petrosian radius is approximately twice the effective radius \citep{Graham...2005}, so the faintest effective surface brightness will be given by
\begin{eqnarray}
\mu_\mathrm{e} &\simeq& 16 + 2.5 \log 2 \pi (6 \, \mathrm{arcsec} / 2)^2 \\
&=& 20.38\,\mathrm{mag\,arcsec}^{-2}
\end{eqnarray}
using the limits in Petrosian magnitude and radius described above. Fainter than this, there is a sharp decrease in the number counts as a function of surface brightness. We can therefore expect to find significant incompleteness at low surface brightness.

\begin{figure}
\centerline {
\includegraphics[width=0.5\textwidth]{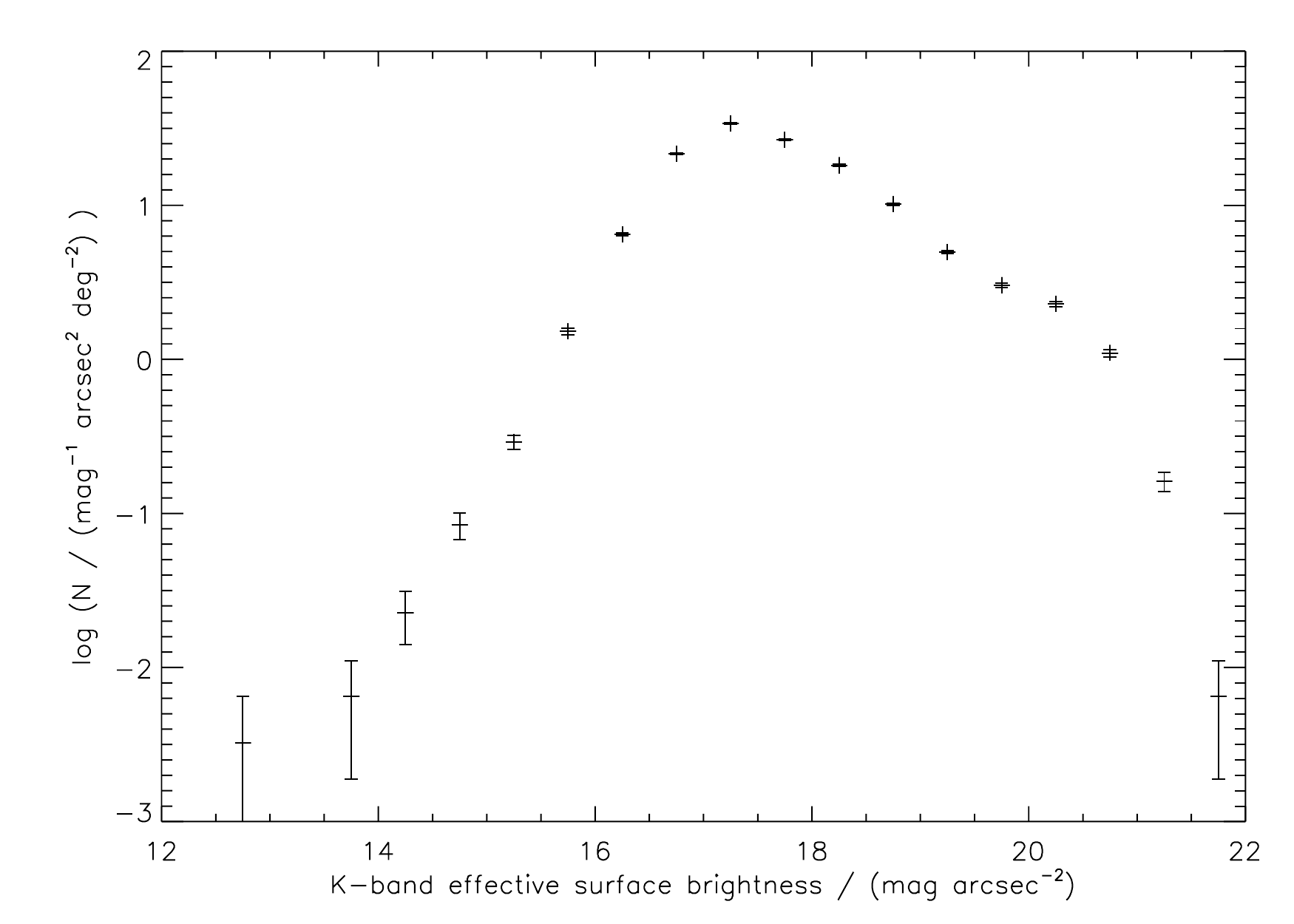}}
\caption[sb_counts]{\label{sb_counts} Number counts as a function of $K$-band effective surface brightness. The limits in magnitude and radius have been imposed on the matched sample, leaving 40\,330 sources. Poisson errors are shown.}
\end{figure}

Fig.\ \ref{sb_counts} shows the number counts as a function of $K$-band effective surface brightness, after the limits in magnitude and radius have been applied.

The SDSS main galaxy sample has a limit of $\mu_\mathrm{e} \leq 24.5$\,mag\,arcsec$^{-2}$ \citep{Strauss...2002}. This limit is taken into account here, although it has a negligible effect on our results. Very few galaxies in our sample have $\mu_{\mathrm{e},r} > 23$\,mag\,arcsec$^{-2}$, so we assume the limit in SDSS surface brightness adds no further incompleteness to the sample, once the magnitude limits in $r$ and $K$ and the surface brightness limit in $K$ have been considered.

\subsection{Redshifts}
\label{sec:redshift}

Galactocentric velocity corrections are applied \citep{Loveday2000}, which typically change the redshifts such that each galaxy in this sample is $0.005 \pm 0.01$\,mag fainter, with some low-redshift galaxies ($z \gtrsim 0.01$) changed by almost 0.1\,mag.

\begin{figure}
\centerline {
\includegraphics[width=0.5\textwidth]{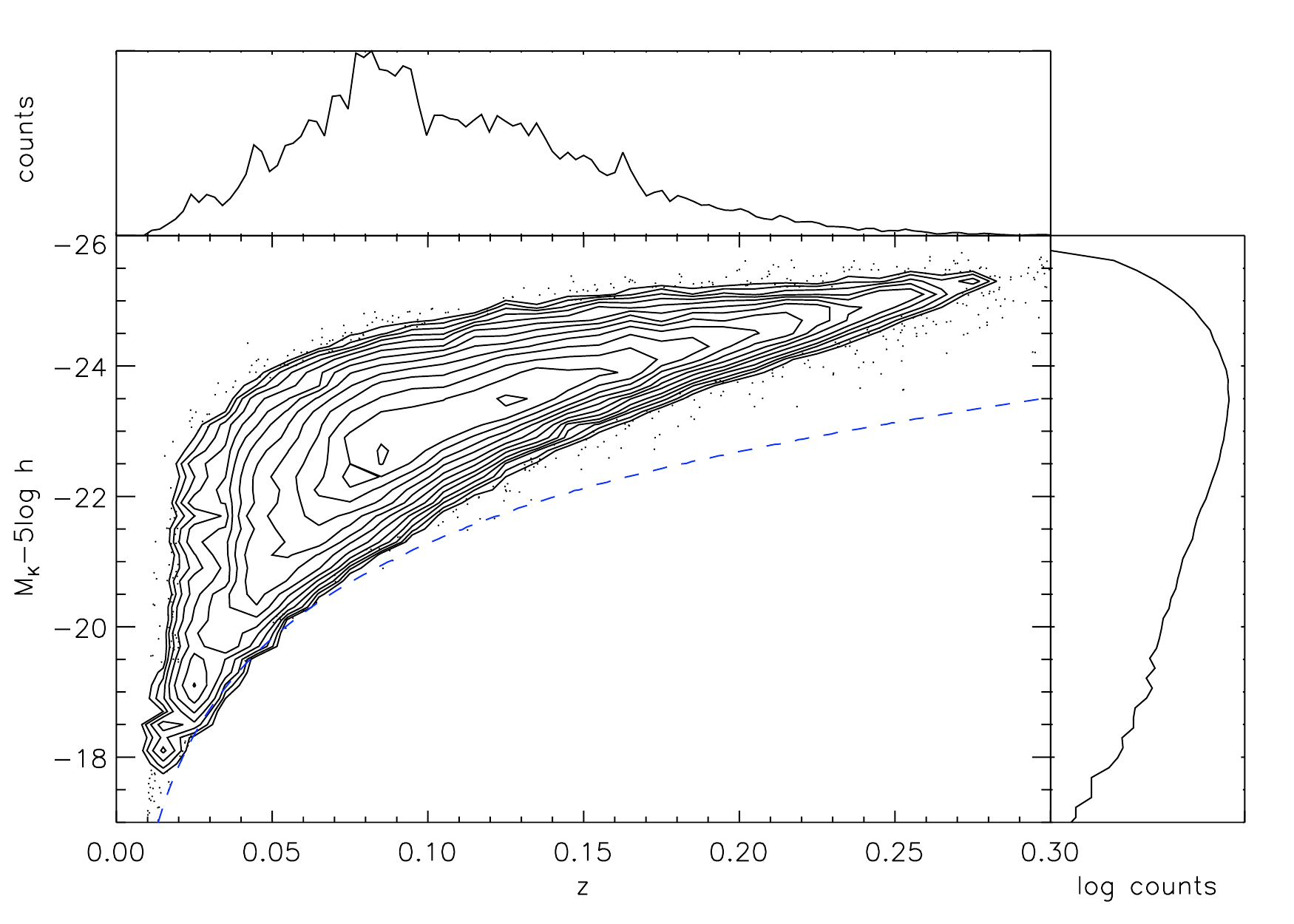}}
\caption[zhist]{\label{zhist} Redshift and $K$-band absolute magnitude distribution of the sample (contours and points), with histogram of redshift distribution and (logarithmic) histogram of $K$-band absolute magnitude distribution. For reference, the absolute magnitude as a function of redshift corresponding to a source at the $K$-band faint magnitude limit, with typical $K$- and evolution-corrections and neglecting the $r$-band limit, is shown by the blue dashed curve. It can be seen that relatively few galaxies are observed near the $K$-band magnitude limit; this is because of the $r$-band magnitude limit. Redshift distributions for a subdivided sample are shown in Fig.\ \ref{lf_ngpsgp}.}
\end{figure}

Fig.\ \ref{zhist} shows the distribution of the sample in redshift and $K$-band absolute magnitude, excluding those sources that lie outside the limits in (apparent) magnitude, radius, surface brightness and redshift used in the analysis below. The presence of large-scale structure can be seen.

A high redshift limit of $z<0.3$ is imposed to limit the effect of $K$- and evolution-corrections (see below).

A low redshift limit of $z>0.01$ is chosen to limit the effect of peculiar velocities, which are not taken into account. Note that this limit would need to be relaxed in order to sample galaxies with very low luminosity.

Table \ref{tbl:limits} shows the various limits on the sample.

\begin{table}
\caption[Sample limits]{\label{tbl:limits} Limits set on observed quantities, used to define the sample and to estimate the contribution of each galaxy to the space density.}
\vspace{.2in}
\centering
\begin{tabular}{lcc}
\hline
\hline
Quantity & Minimum & Maximum \\
\hline
$K$ Petrosian magnitude & - & 16\,mag \\
$r$ Petrosian magnitude & - & 17.6\,mag \\
$g$ fiber magnitude & 15\,mag & - \\
$i$ fiber magnitude & 14.5\,mag & - \\
$K$ Petrosian radius & - & 6\,arcsec\\
$\mu_{\mathrm{e},K}$ & - & 21\,mag\,arcsec$^{-2}$ \\
$\mu_{\mathrm{e},r}$ & - & 24.5\,mag\,arcsec$^{-2}$ \\
$z$ & 0.01 & 0.3\\
\hline
\end{tabular}
\end{table}

\subsection{$K$- and evolution-corrections}

$K$-corrections (to $z=0$) are estimated using \textsc{kcorrect} v4\_1\_4 \citep{BlantonR2007}. The five optical Petrosian magnitudes ($ugriz$), before the evolution corrections have been applied (see below), are used to fit galaxy templates to each galaxy, from which the $K$-correction is derived in $r$ and $K$. WFCAM filter files have been generated from \citet{Hewett...2006}. Due to the inconsistent Petrosian apertures between UKIDSS and SDSS bands, we do not have good optical--NIR galaxy colours and so we have not been able to use the full set of bands ($ugrizYJHK$) for the template fitting.

When covering a significant range in redshift, it is important to include evolution corrections.  This is because the bright end of the LF will consist mainly of galaxies at high redshift, while the faint end will be made up from galaxies at low redshift. A failure to include evolution corrections leads to a distortion in the shape of the luminosity function.

Evolution corrections are applied using a simple $E(z) = Qz$, where $Q$ in $ugriz$ is taken from \citet{Blanton...2003c} and we take $Q=1$ in $K$ \citep[consistent with stellar population synthesis models,][Section 6.2]{Blanton...2003c}, i.e.\ $Q = (4.22, 2.04, 1.62, 1.61, 0.76, 1.0)$ for $ugrizK$. Note that these evolution corrections are very simplistic, since different kinds of galaxies evolve in different ways, but the correction is small: at $z=0.3$, the $K$-band evolution correction is $Qz=0.3$\,mag.

Absolute magnitudes are given by
\begin{eqnarray}
M = m - DM(z) - K(z) + E(z)
\end{eqnarray}
where $DM(z)$ is the distance modulus and $K(z)$ is the $K$-correction. Absolute surface brightness, in mag\,arcsec$^{-2}$, is given by
\begin{eqnarray}
\mu_\mathrm{abs} = \mu - 10 \log(1+z) - K(z) + E(z)\,.
\end{eqnarray}
Note that in the $K$ band, for all galaxies in the sample, the absolute surface brightness is not more than 0.5\,mag brighter than the apparent surface brightness. $K$- and evolution-corrections are not applied when estimating the physical radius, which is given in $h^{-1}$\,kpc as
\begin{eqnarray}
R = \frac{1000\pi \theta D_\mathrm{A}}{ 180 \times 3600}
\end{eqnarray}
where $\theta$ is the angular size in arcsec and $D_\mathrm{A}$ is the angular diameter distance in $h^{-1}$\,Mpc.

\begin{figure}
\centerline {
\includegraphics[width=0.5\textwidth]{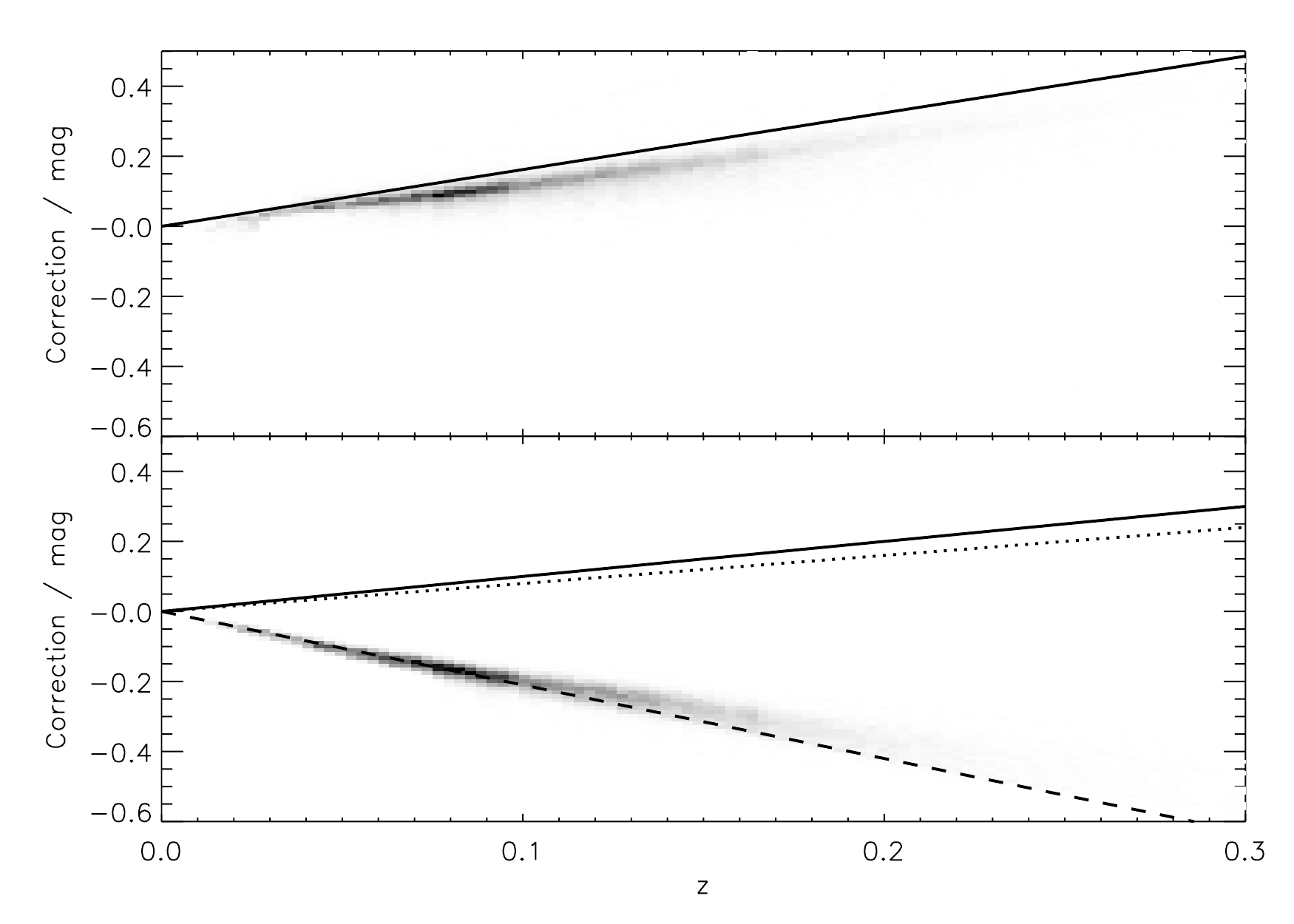}}
\caption[kecorr]{\label{kecorr} $K$- (shaded regions) and evolution- (solid lines) corrections in the $r$ band (upper panel) and in the $K$ band (lower panel). Also shown for the $K$ band are the mean $K$-correction (dashed line, $K(z) = -2.1z$) and evolution-correction (dotted line, $E(z) = 0.8z$) from \citet{Bell...2003c}. (These are simple approximations, so close agreement is not to be expected.)}
\end{figure}

Fig.\ \ref{kecorr} shows the $K$- and evolution-corrections in the $K$ band and $r$ band, also showing the $K$-band corrections used by \citet{Bell...2003c}. Their $K$-corrections are stronger than those used here, but their value of $Q=0.8$ is weaker than our value.  These two largely cancel each other out, with our $K(z) - E(z)$ being approximately the same as theirs for low redshift.

\subsection{Dust}

\citet{Driver...2007b} have shown that dust has a considerable effect on the shape of the optical LF, and caution that the effect could still be significant in the NIR. To correct for this inclination-dependent dust attenuation would ideally require good bulge-to-disc decomposition, which is beyond the scope of this work. We choose here to present results that are not corrected for dust, thus representing the amount of $K$-band light that escapes from the galaxy, rather than the amount of $K$-band light emitted by the stars in the galaxy (some of which will be absorbed by dust).  However, in estimating the absolute magnitude of a galaxy we have implicitly assumed isotropic emission of light, which is not the case: a disc galaxy will appear fainter when viewed edge-on than when viewed face-on, leading to a corresponding under- or over-estimate, respectively, of the galaxy's total (attenuated) luminosity. This will lead to a blurring of the LF for disc galaxies, but less so than at optical wavelengths.

\section{Method}

In a survey limited by more than one measured quantity, e.g.\ two magnitudes, or magnitude and surface brightness, there are two ways to proceed.  One is explicitly to take into account all of the known selection effects. The other is to set strong limits in one or more quantities so that the selection effects in the remaining quantities are negligible. Either of these ensures that completeness is taken into account and that the contribution of each observed galaxy to the total space density is estimated accurately (by $1/V_\mathrm{max}$).

There are three traditional methods for estimating the LF. If the functional form of the LF is known, the best method to use is the STY \citep*{SandageTY1979} maximum likelihood method.  But in this work the space density is a four-dimensional distribution and the functional form is unknown, so the STY method cannot be used. The other two methods have no prior assumptions about the functional form. These are the stepwise maximum likelihood (SWML) method \citep*{EfstathiouEP1988}, a version of the STY method where the functional form is described by a value in each bin, and the $1/V_\mathrm{max}$ method \citep{Schmidt1968}, which is a simple estimator of the value of the space density in each bin. The SWML method works by factoring out the normalization at each redshift and matching the shape of the LF across different redshifts.  It is therefore less prone to clustering effects than the $1/V_\mathrm{max}$ method, which assumes a homogeneous distribution, but it (like the STY method) does not give a value for the overall normalization, which must be determined by other means. However, the bins in the SWML method are strongly correlated, so any problem in the SWML estimation is likely to affect the entire luminosity function. The $1/V_\mathrm{max}$ estimator is more robust against systemic failure, but requires a large survey volume in order to overcome the effects of large-scale structure.

In this work, due to the complexity of the four-dimensional space, we have found the SWML method to be prone to fail, whereas the $1/V_\mathrm{max}$ method has proved to be more stable.  Large-scale structure has not been taken into account, but a large volume has been probed at bright luminosities and faint luminosities are already affected by incompleteness at low surface brightness.

The volume within which each galaxy would be visible, $V_\mathrm{max}$, is estimated for each galaxy by considering the tightest constraints on maximum and minimum redshift provided by the limits in $z$, $K$, $r$, $K$-band $\mu_\mathrm{e}$ and $K$-band Petrosian radius. Each galaxy is then given a weight $w_i = 1 / V_{\mathrm{max},i}$.  The space density $\phi$ of galaxies of a certain type (e.g., binned in absolute magnitude and/or surface brightness) is given by $\phi = \sum_i w_i$ for galaxies $i$ of that type.

The effect of the $1 / V_\mathrm{max}$ factors on the features of the $K$-band LF may be appreciated by comparing the shape of the absolute magnitude histogram in Fig.\ \ref{zhist} with the shape of the LF in Fig.\ \ref{lf}. For galaxies with higher luminosity, $V_\mathrm{max}$ is considerably greater, so the effect of the weighting on the shape of the curve is to raise the faint end relative to the bright end.  It may also be seen that the $1/V_\mathrm{max}$ weighting introduces no obviously spurious features into the final LF, despite the complexity of the limits on the sample.

Some regions of the four-dimensional parameter space have not been sampled, for example, low-surface brightness galaxies, so care needs to be taken when interpreting the binned space density.

\subsection{Errors}
\label{jackknife}

Statistical errors for all quantities (except where indicated) are estimated through the jackknife method \citep{Lupton1993}.  The sample area is divided into 24 regions, each containing approximately the same number of galaxies; the principal subdivisions are shown in Fig.\ \ref{coverage}. The space density is calculated once for the whole sample and then a further 24 times, each time omitting one of the 24 regions. The variance of each value of $\phi$ is then calculated using:
\begin{eqnarray}
\mathrm{Var} (\phi) = \frac{n-1}{n} \sum_i (\phi^\mathrm{J}_i - \overline{\phi^\mathrm{J}} )^2
\end{eqnarray}
where $\phi^\mathrm{J}_i$ is the $i$th jackknife resampling of the data and $n$ is the number of jackknife resamplings (24 in this case). A bias correction \citep{Lupton1993} is applied to the space density, giving a new value of
\begin{eqnarray}
\phi' = \phi + (n - 1) (\phi - \overline{\phi^\mathrm{J}})
\end{eqnarray}
where $\phi$ is the original estimate for the whole sample.

Uncertainties and bias corrections for the functional fits and integrated quantities (for example, luminosity density) are estimated using the same method.

It would be preferable, but more difficult, to divide the sample into regions of equal area rather than equal numbers of galaxies. Given the way in which the effective area is calculated, these subdivisions cannot say anything about the overall normalization of the space density; for example, whether the south Galactic pole has a lower density of galaxies than the north Galactic pole.

Uncertainties in the magnitudes and other galaxy properties are not included in the analysis. This is likely to have only a small effect on most of our results, but the magnitude errors will cause us slightly to overestimate the space density of galaxies at the bright end of the luminosity function, where the LF is very steep \citep{Jones...2006}.

\subsection{Functional fits}

Functional fits for the BBD, LFs and SMF are found using the \textsc{mpfit} routines written by Craig Markwardt.\footnote{\url{http://cow.physics.wisc.edu/~craigm/idl/fitting.html}}

For the LFs and the SMF, the Schechter function is used. Expressed in magnitudes, this takes the form
\begin{eqnarray}
\phi(M) = 0.4 \ln 10 \phi^* 10^{0.4(M^*-M)(\alpha + 1)} \mathrm{e}^{-10^{0.4(M^*-M)}}
\end{eqnarray}
where $M^*$ is the characteristic absolute magnitude, $\alpha$ is the faint end slope and $\phi^*$ is the normalization.

For the BBD, the \citet{Choloniewski1985b} function is used \citep{Driver...2005}:
\begin{eqnarray}
\begin{split}
\phi(M, \mu_\mathrm{e}) =& \frac{0.4 \ln 10}{\sqrt{2\pi} \sigma_{\mu_\mathrm{e}}} \phi^* 10^{0.4(M^*-M)(\alpha + 1)} \mathrm{e}^{-10^{0.4(M^*-M)}} \\
& \times \exp \left\{ -\frac{1}{2} \left[ \frac{\mu_\mathrm{e} - \mu^*_\mathrm{e} - \beta (M - M^*)}{\sigma_{\mu_\mathrm{e}}} \right]^2 \right\}
\end{split}
\end{eqnarray}
where $M^*$, $\alpha$ and $\phi^*$ are the usual Schechter function parameters, $\mu^*_\mathrm{e}$ is the mean surface brightness at $M^*$, $\sigma_{\mu_\mathrm{e}}$ is the standard deviation in surface brightness and $\beta$ is the slope of the relationship between absolute magnitude and mean surface brightness. Note that integrating the Cho\l oniewski function over surface brightness gives a Schechter function.

The range of points over which the functions are fit is restricted in order to avoid known regions of incompleteness: only data brighter than $-20$ in $M_K - 5 \log h$, $-18$ in $M_r - 5 \log h$ and 19 in surface brightness, or more massive than 9.5 in $\log (h^{-2} $M$_\odot)$, are used to generate the functional fits.

\subsection{Testing the method with simulated data}

We test the $V_\mathrm{max}$ estimator and fitting routines using four simulated samples drawn from a known Cho\l oniewski function, with Gaussian $r-K$ colours (mean and standard deviation derived from the observed sample, but with no dependence on luminosity), and subject to the same observational limits of the data sample, except with $r < 17.77$ rather than $r < 17.6$. Each sample contains between 42\,632 and 42\,885 galaxies.

\begin{figure}
\centerline {
\includegraphics[width=0.5\textwidth]{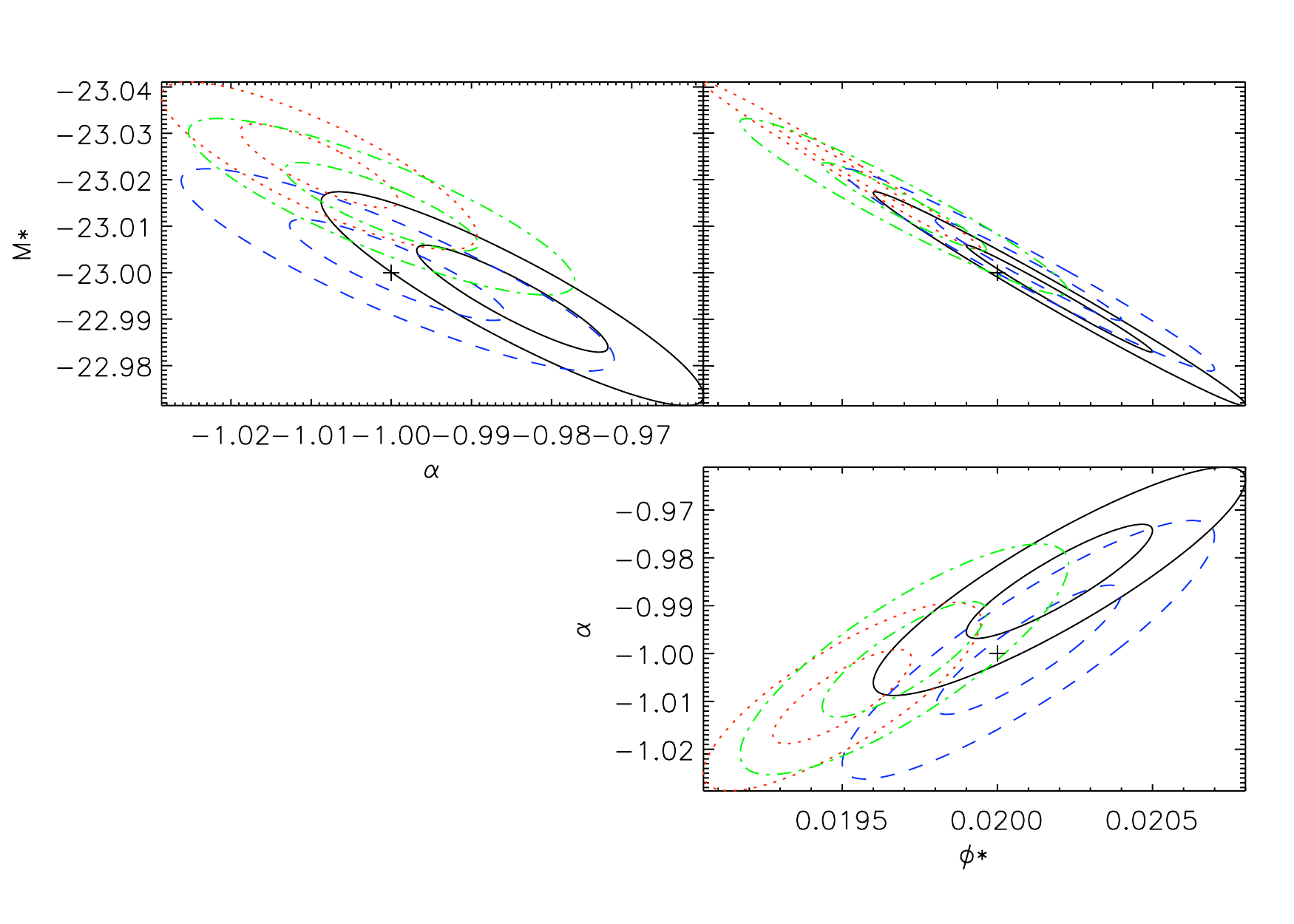}}
\caption[sch_fit_params]{\label{sch_fit_params} Schechter function parameters for the simulated samples. Plus symbols show the input parameters, $M^*-5\log h=-23$, $\alpha=-1$ and $\phi^*=0.02h^3$\,Mpc$^{-3}$, while contours show the $1\sigma$ and $2\sigma$ error contours on the recovered parameters for the four different simulated samples. These uncertainties are estimated from 24 jackknife estimations of the Schechter function.}
\end{figure}

Fig.\ \ref{sch_fit_params} shows the recovered Schechter function parameters from the simulated samples.  There is no obvious systematic bias. The recovered uncertainties, shown by the contours, give an approximate measure of the closeness of the recovered parameters to the input Schechter function, although they tend to underestimate the true error.

The luminosity density, in solar units, may be calculated from the Schechter function as
\begin{eqnarray}
j = \int_0^\infty L \phi(L) \,\d L = \phi^* 10^{0.4 (M_\odot - M^*)} \Gamma(\alpha + 2)
\end{eqnarray}
where $M_\odot$ is the absolute magnitude of the sun \citep[taken to be 3.32 in the $K$-band,][]{Bell...2003c}, or from the weights of each galaxy by
\begin{eqnarray}
j = \sum_i 10^{0.4 (M_\odot - M_i)} w_i \,.
\end{eqnarray}
For the simulated samples the luminosity density was estimated in these two ways and compared to the luminosity density from the input Schechter function.  By summing the galaxy weights, the recovered luminosity density typically underestimated the input luminosity density by around 1 per cent, whereas there was no obvious bias from integration of the recovered Schechter function.

\begin{figure}
\centerline {
\includegraphics[width=0.5\textwidth]{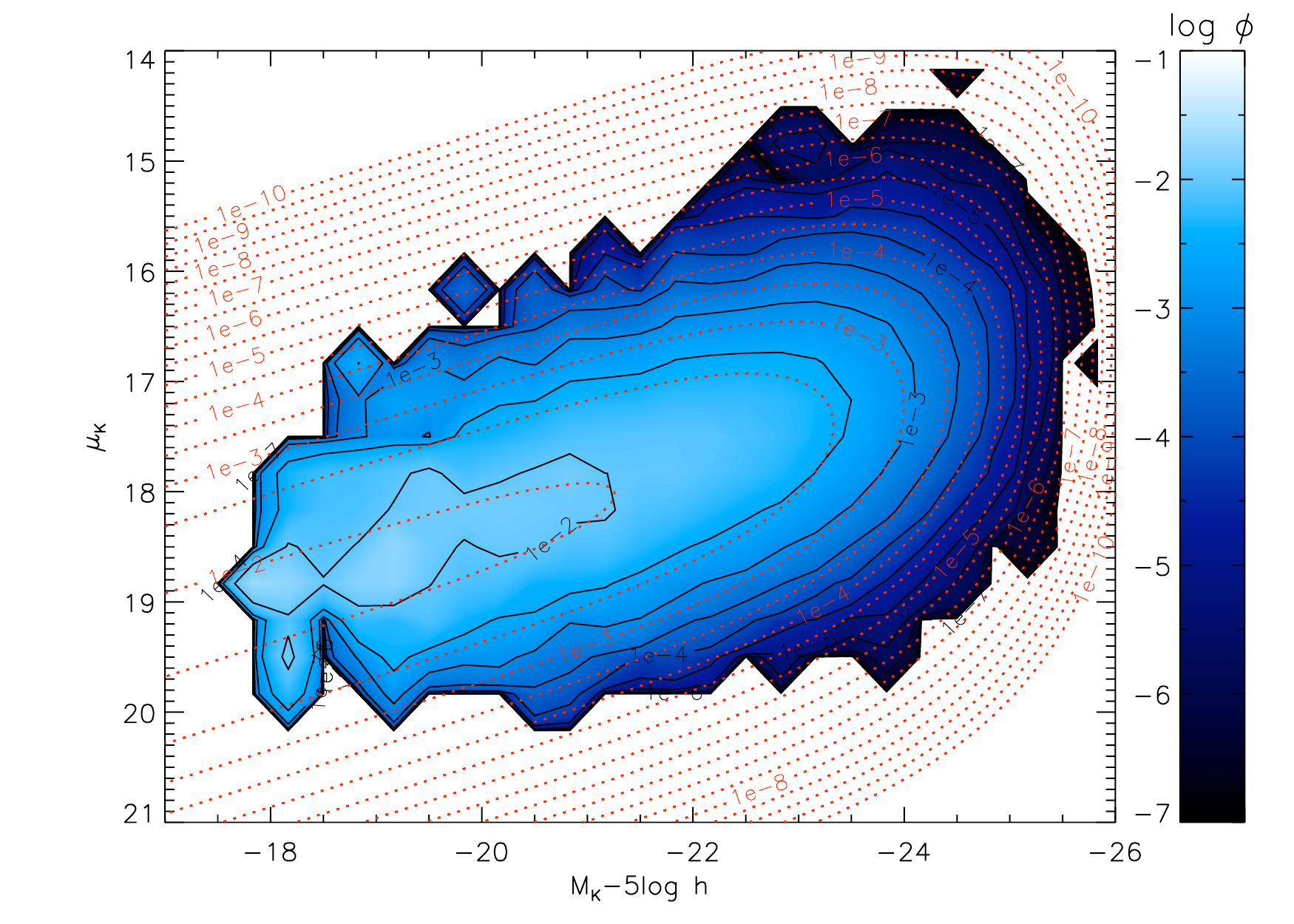}}
\caption[chol_fit_bbd]{\label{chol_fit_bbd} Input Cho\l oniewski function for the simulations (red dotted contours) and the recovered BBD from one of the simulated samples (shaded regions and black solid contours). The space density, $\phi$, is in units of $h^3$\,Mpc$^{-3}$\,mag$^{-1}$\,(mag\,arcsec$^{-2}$)$^{-1}$. A Cho\l oniewski function was fit to the recovered BBD (not shown). The input (recovered) Cho\l oniewski function parameters are $M^*-5\log h=-23(-23.03 \pm 0.01)$\,mag, $\alpha=-1(-0.96 \pm 0.01)$,  $\phi^*=0.02(0.0224 \pm 0.0003)h^3$\,Mpc$^{-3}$, $\mu^*_\mathrm{e}=17.5(17.363 \pm 0.005)$\,mag\,arcsec$^{-2}$, $\sigma_{\mu_\mathrm{e}}=0.6(0.580 \pm 0.003)$\,mag\,arcsec$^{-2}$ and $\beta=0.3(0.287 \pm 0.003)$. Very similar results are obtained for the other three simulated samples.}
\end{figure}

Fig.\ \ref{chol_fit_bbd} shows the input Cho\l oniewski function and the recovered BBD for one of the simulated samples. The Cho\l oniewski function gives a good fit to the recovered BBD, but the recovered BBD is itself biased with respect to the input Cho\l oniewski function, for all four simulated samples, most noticeably towards higher surface brightness, by approximately 0.15\,mag\,arcsec$^{-2}$. The BBD presented below should therefore be considered only approximately correct.

\section{Bivariate brightness distribution and luminosity functions}

In this section the bivariate brightness distribution and luminosity functions are calculated for the whole sample and for subdivisions of the data.

\subsection{Bivariate brightness distribution}
\label{sec:bbd}

\begin{figure}
\centerline {
\includegraphics[width=0.5\textwidth]{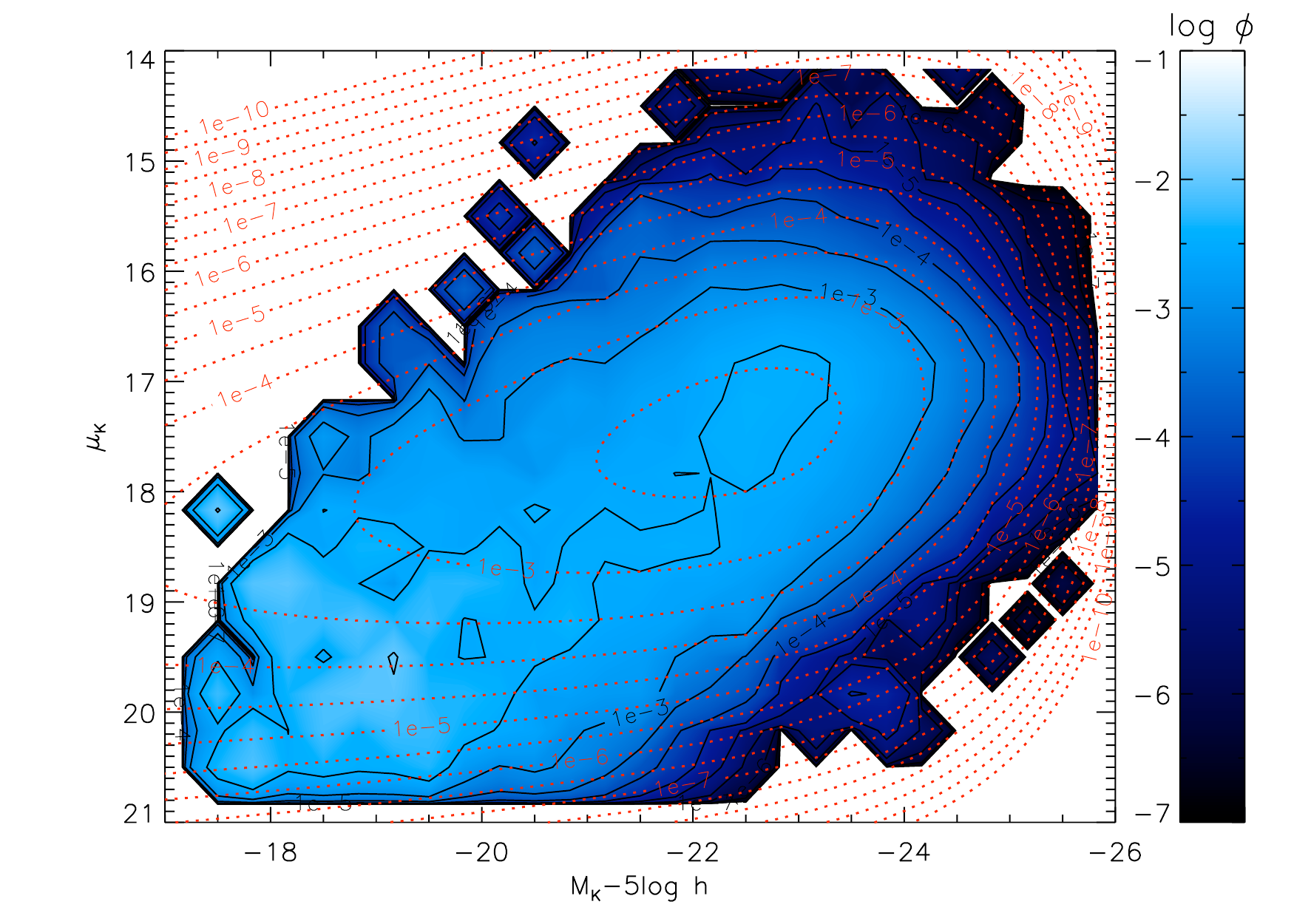}}
\caption[bbd]{\label{bbd} BBD for the full sample in $K$-band absolute magnitude and absolute effective surface brightness.  Shaded regions and solid black contours show the space density, $\phi$, as in Fig.\ \ref{chol_fit_bbd}.  The best-fitting Cho\l oniewski function, estimated using $M_K - 5 \log h < -20$ and $\mu_\mathrm{e,abs} < 19$, is shown by the red dotted contours. Parameters of the fit are $M^*-5\log h=-22.96$\,mag, $\alpha=-0.38$,  $\phi^*=0.0201h^3$\,Mpc$^{-3}$, $\mu^*_\mathrm{e,abs}=17.36$\,mag\,arcsec$^{-2}$, $\sigma_{\mu_\mathrm{e,abs}}=0.672$\,mag\,arcsec$^{-2}$ and $\beta=0.188$.}
\end{figure}

Fig.\ \ref{bbd} shows the BBD in $K$-band absolute magnitude and absolute effective surface brightness. The value of the space density at any point on the BBD is estimated assuming the visibility of the full range of galaxy types that exist with that absolute magnitude and surface brightness.  This may in fact not be the case, given the additional limits in faint $r$-band apparent magnitude and large $K$-band Petrosian radius.

\begin{figure}
\centerline {
\includegraphics[width=0.5\textwidth]{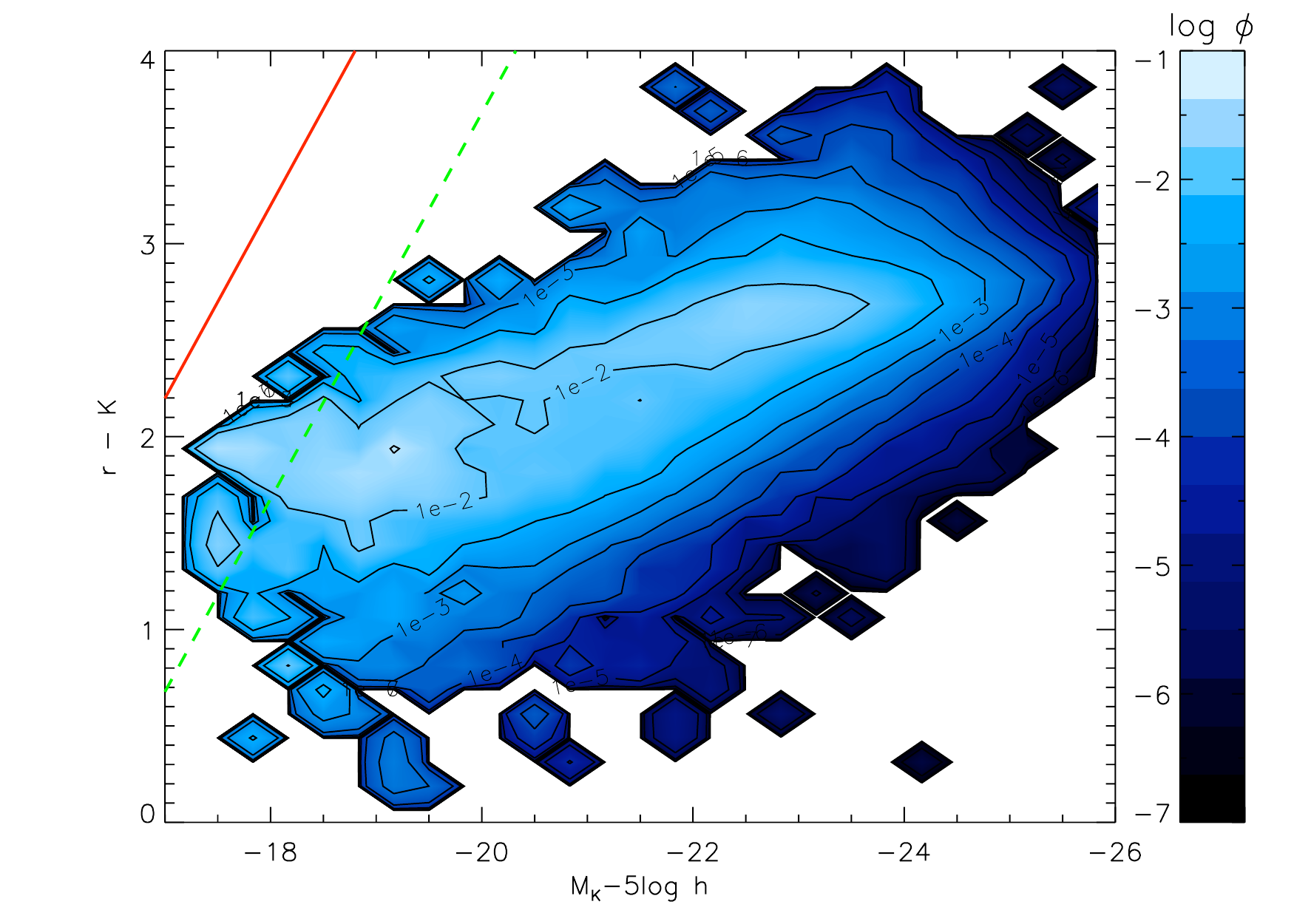}}
\caption[rmk_k]{\label{rmk_k} Space density of galaxies, in units of $h^3$\,Mpc$^{-3}$\,mag$^{-2}$, as a function of $K$-band absolute magnitude and rest-frame $r-K$ Petrosian colour.  Note that this is not a true colour since the apertures differ between the bands. The straight lines show the position on the plot of hypothetical galaxies at the $r$-band flux limit, with various $r-K$ colours, situated at $z=0.01$ (upper-left line, solid, red) or at $z=0.02$ (lower-right line, dashed, green). $z=0.02$ corresponds to a survey volume of $1.17 \times 10^4h^{-3}$\,Mpc$^3$.}
\end{figure}

\begin{figure}
\centerline {
\includegraphics[width=0.5\textwidth]{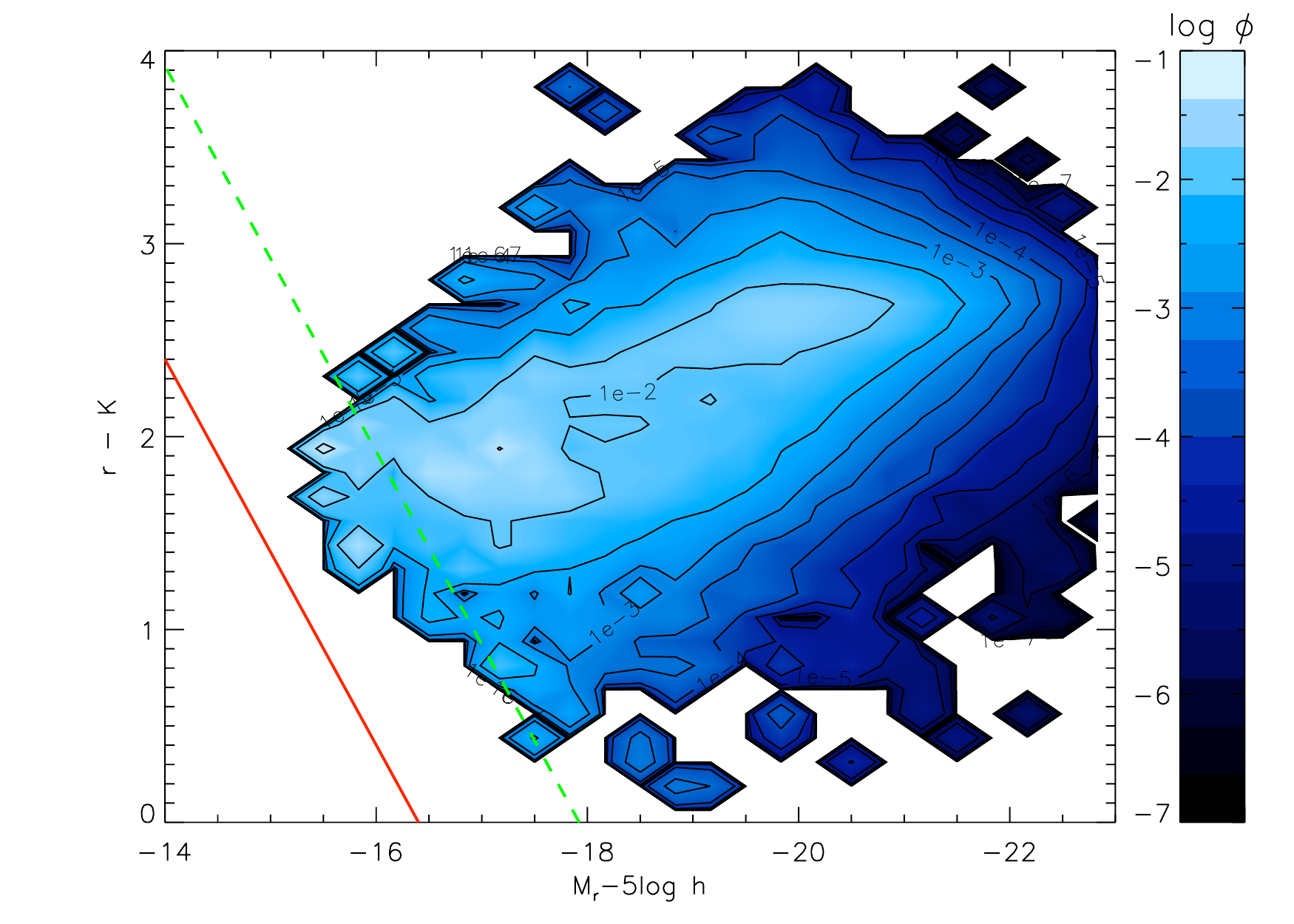}}
\caption[rmk_r]{\label{rmk_r} As Fig.\ \ref{rmk_k} but for $r$-band absolute magnitude. The straight lines correspond to hypothetical galaxies at the $K$-band flux limit, with various $r-K$ colours, situated at $z=0.01$ (lower-left line, solid, red) or at $z=0.02$ (upper-right line, dashed, green).}
\end{figure}

Figs.\ \ref{rmk_k} and \ref{rmk_r} show the effect of the combined $r$- and $K$-band flux limits on the completeness, as a function of $K$- and $r$-band absolute magnitude respectively.

From Fig.\ \ref{rmk} it can be seen that near the $r$-band flux limit blue galaxies cannot be seen, and that near the $K$-band flux limit red galaxies cannot be seen. For the red galaxies that would not be seen at near the $K$-band flux limit, we consider galaxies at redshift $z$ with $r=17.6$ and various $r-K$ colours, neglecting $K$- and evolution-corrections. These sources will have $K$-band absolute magnitude
\begin{eqnarray}
M_K = K - DM(z) = r - (r - K) - DM(z)\,.
\end{eqnarray}

Fig.\ \ref{rmk_k} shows this relation between $M_K$ and $r-K$ (with $r=17.6$) for $z=0.01$ and 0.02, corresponding respectively to the minimum redshift considered here and the redshift at which the survey volume is just over $10^4h^{-3}$\,Mpc$^3$. It can be seen that red galaxies will not be seen at faint $K$-band luminosity, and we therefore anticipate some incompleteness at $M_K - 5 \log h > -19$.

A similar relation may be derived for the blue galaxies that would not be seen near the $K$-band flux limit.  Considering a galaxy at redshift $z$ with $K=16$ (the $K$-band flux limit), the $r$-band absolute magnitude will be
\begin{eqnarray}
M_r = r - DM(z) = K + (r - K) - DM(z)\,.
\end{eqnarray}

Fig.\ \ref{rmk_r} shows this relation between $M_r$ and $r-K$ (with $K=16$), from which it may be seen that blue galaxies will not be seen at faint $r$-band luminosity. Significant incompleteness is to be expected for $M_r - 5 \log h > -17.5$.

For the large radius limit, considering a source with a certain $K$-band magnitude and surface brightness, and assuming the Petrosian radius $r_\mathrm{P}$ is twice the effective radius, Equation (\ref{sb}) can be written as
\begin{eqnarray}
\mu_\mathrm{e} \simeq K + 2.5 \log 2 \pi \left( \frac{r_\mathrm{P}}{2} \right)^2\,.
\end{eqnarray}
For $K$=16 and $\mu_\mathrm{e}=19.5$\,mag\,arcsec$^{-2}$, this corresponds to a limit in Petrosian radius of 4.0 arcsec, considerably less than the 6 arcsec limit intrinsic to the data; only for surface brightness fainter than 20.38\,mag\,arcsec$^{-2}$ does the 6 arcsec large radius limit dominate.

Fig.\ \ref{bbd} also shows the best-fitting Cho\l oniewski fit to the BBD.  It can be seen that the function provides a poor fit to the data, being unable to fit simultaneously the decline at high surface brightness and high luminosity, the broadening of the surface brightness distribution at faint luminosity, or the slope of the luminosity--surface brightness relation, which flattens at high luminosity. These features have also been seen in optical determinations of the BBD \citep[e.g.,][]{Driver1999, Blanton...2001, CrossD2002, Driver...2005}

\subsection{$K$-band LF and luminosity density}

\begin{figure*}
\centerline {
\includegraphics[width=0.8\textwidth]{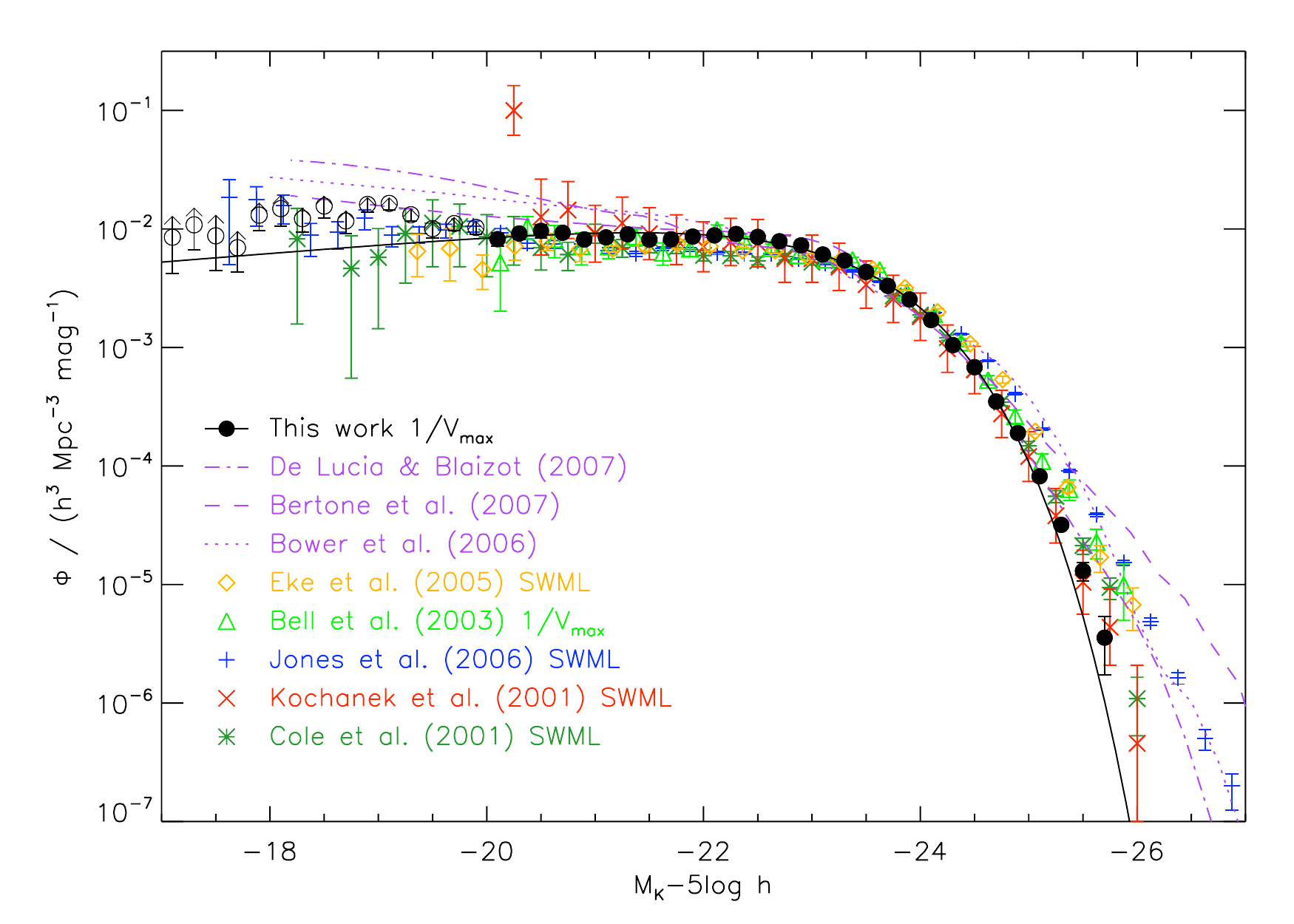}}
\caption[lf]{\label{lf} $K$-band LF for the whole sample, with a compendium of published results from observations or semi-analytic models. Only the filled points are used in the Schechter function fit, i.e., $M_K - 5\log h < -20$; the unfilled points are likely to suffer from some incompleteness of low-surface brightness or red, low-luminosity galaxies. Schechter function parameters are $M^*-5\log h=-23.19 \pm 0.04$, $\alpha=-0.81 \pm 0.04$ and $\phi^*=(0.0166 \pm 0.0008)h^3$\,Mpc$^{-3}$, although, as noted in the main text, the Schechter function provides a poor fit to the LF at both high and low luminosity. \nocite{deLuciaB2007,BertonedLT2007,Bower...2006,Eke...2005,Bell...2003c,Jones...2006,Kochanek...2001b}}
\end{figure*}

Fig.\ \ref{lf} shows the $K$-band LF for the whole sample.  The parameters of the best-fitting Schechter function correlate strongly, with corr$(M^*,\alpha)=0.92$, corr$(M^*,\phi^*)=0.97$ and corr$(\alpha,\phi^*)=0.92$.

The most significant deviation from the published $K$-band luminosity functions is at the bright end, where our LF has a very steep decline compared with those of \citet{Bell...2003c}, \citet{Eke...2005} and \citet{Jones...2006}. There are several possible explanations for this. For example, the following explanations are given.
\begin{enumerate}
\item Differences in the evolution corrections used (0.3\,mag at $z=0.3$ for our value of $Q=1$), affecting the high-luminosity galaxies (cf.\ Fig.\ \ref{zhist}). \citet{Jones...2006} applied no evolution corrections, although they have a redshift limit of $z<0.2$. \citet{Eke...2005}, for $z<0.12$, and \citet{Bell...2003c} applied combined $K$- and evolution-corrections similar to our own.  There is better agreement between our results and those of \citet{Kochanek...2001b}, although they applied no evolution corrections.
\item We have used Petrosian magnitudes, which are significantly fainter than total magnitudes for galaxies with a high S\'ersic index \citep{Graham...2005}, for example, by 0.22\,mag for a de Vaucouleurs profile \citep{Blanton...2001}. \citet{Kochanek...2001b} used isophotal magnitudes with a correction of $0.20 \pm 0.04$\,mag to estimate total magnitudes, but a larger correction may be needed for the most luminous galaxies \citep{Blanton...2001}, which tend to have a higher S\'ersic index. \citet{Bell...2003c} used the 2MASS Kron magnitudes with a correction of 0.1\,mag, \citet{Eke...2005} applied a similar correction based on the $J-K_\mathrm{S}$ colour, while \citet{Jones...2006} used the 2MASS total (extrapolated) magnitudes. However, this effect is countered to an extent by the effects of seeing, which, for a de Vaucouleurs profile, increases the fraction of the galaxy's flux recovered by the Petrosian magnitude when the angular size of the galaxy is small \citep{Blanton...2001}. This is the case for most of the luminous galaxies in the sample, which are generally observed at higher redshift.
\item Unidentified sources of incompleteness or a poorly understood selection function, given the non-trivial limits on our sample and the dependence on completeness in SDSS.
\item Improved photometry: significant magnitude errors at the bright end will lead to an overestimate of the space density at high luminosities \citep{Jones...2006}. The results of \citet{Kochanek...2001b}, \citet{Bell...2003c}, \citet{Eke...2005} and \citet{Jones...2006} are all based on 2MASS magnitudes, with much shallower imaging than the UKIDSS LAS.  Moreover, the most luminous galaxies in our sample are all observed at magnitudes much brighter than the $K$-band flux limit, as can be seen from Fig.\ \ref{zhist}, so the photometric errors are very small. 

Fig.\ \ref{2mass} shows a comparison between $K$-band 2MASS Kron magnitudes and UKIDSS Petrosian magnitudes for sources in this sample with counterparts in the 2MASS extended source catalogue (XSC) \citep{Jarrett...2000a}. It can be seen that a significant number of faint 2MASS sources have 2MASS Kron magnitudes that are much brighter than UKIDSS Petrosian magnitudes, sometimes by over 0.5\,mag. If the UKIDSS magnitudes are more accurate, then this suggests that 2MASS luminosity functions will over-estimate the bright end, since high luminosity galaxies are most often found near the faint magnitude limit. For a more in-depth comparison between UKIDSS and 2MASS photometry, see Cross et al.\ (in preparation).

\begin{figure}
\centerline {
\includegraphics[width=0.5\textwidth]{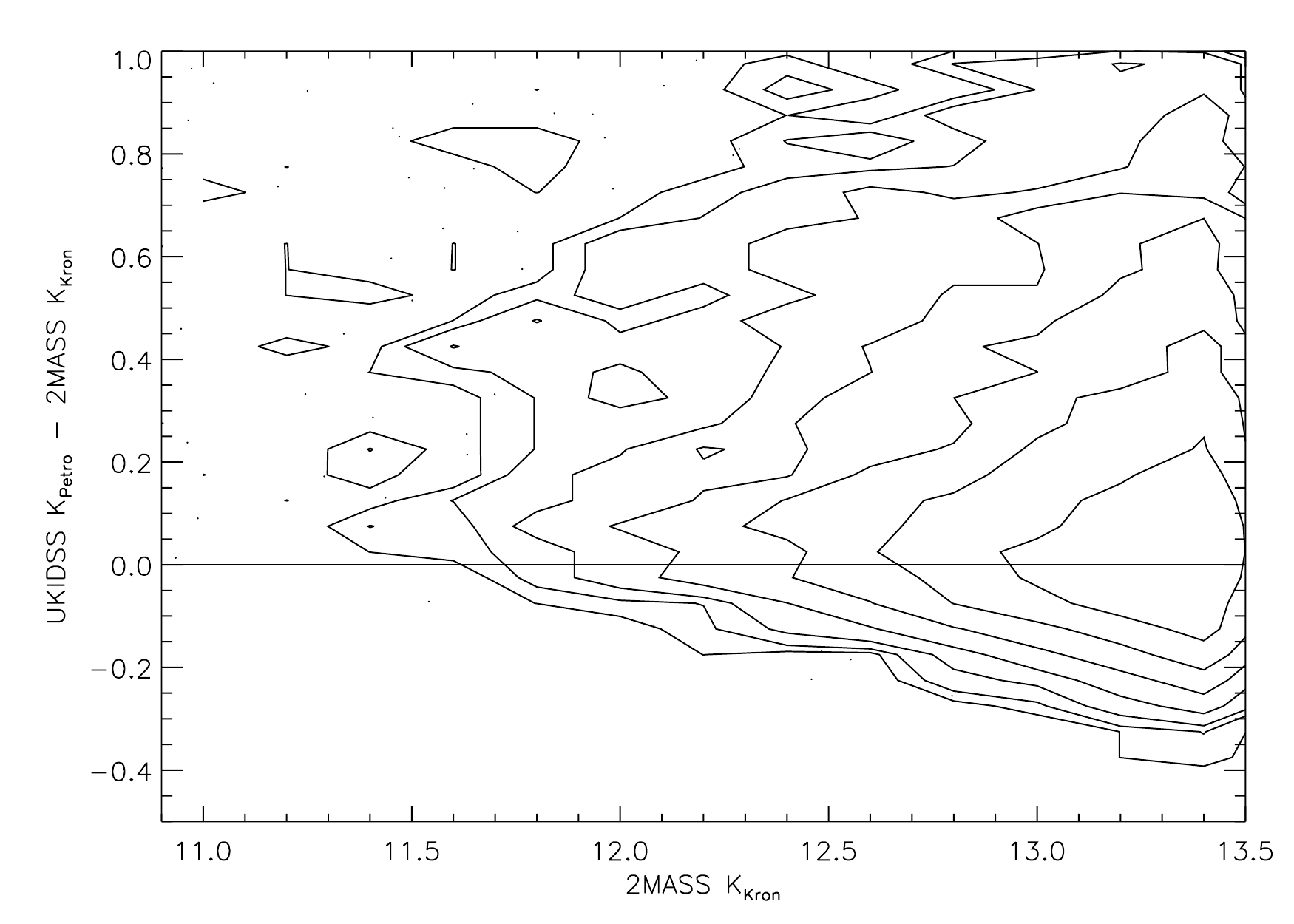}}
\caption[2mass]{\label{2mass} Residual $K$-band magnitudes between 2MASS Kron and UKIDSS Petrosian magnitudes for 6898 galaxies matched between our sample and the 2MASS XSC, with 2MASS $K_\mathrm{Kron}<13.5$\,mag. Dust reddening corrections have not been applied. Contours show the density of sources on a logarithmic scale, with sources shown as points where the density is low. The mean offset is 0.19\,mag, such that 2MASS magnitudes are brighter on average, with a standard deviation of 0.29\,mag.}
\end{figure}

\item Missing sources in high-density environments.  As discussed in Section \ref{sec:area}, the positioning of the SDSS spectroscopic fibres leads to a slight incompleteness at high-density regions. These regions are rich in high-luminosity galaxies, so this may affect the high-luminosity end of the LF.
\item Incompleteness at very large physical radius. As shown in Section \ref{sec:radii}, the limits in $K$-band Petrosian radius and redshift combine to set an upper limit to the physical radius of galaxies in the sample.  It appears likely that this will lead to some incompleteness at high luminosities.
\end{enumerate}
There is general agreement at the bright end of the (Petrosian) $r$-band LF (see below), which is consistent with the possible explanations listed above, although it does suggest that the effect of missing galaxies is not the most significant factor, whether these are large galaxies or galaxies found in high-density environments.  Further investigation is required to resolve this.

The steeper bright end leads to a fainter value of $M^*$ compared with previous results, and consequently (given the strong correlation between $M^*$ and $\alpha$) a shallower value for the faint-end slope, $\alpha$. It should be clear from Fig.\ \ref{lf} that the faint end of the Schechter function does not coincide with the faint end of the LF\@. It should also be noted that the $1/V_\mathrm{max}$ method is sensitive to large-scale structure, which will be particularly apparent at the faint end of the LF.

At intermediate magnitudes, $M_K - 5 \log h \simeq -22$, our LF is noticeably higher than those from the literature. This is likely to be due to an overdensity of galaxies at $z \lesssim 0.1$, as seen in Fig.\ \ref{zhist}. The effect of large-scale structure is discussed in Section \ref{lss}.

The $K$-band luminosity density, with jackknife errors, is found to be $j = (6.185 \pm 0.072) \times 10^8\,$L$_\odot\,h$\,Mpc$^{-3}$ from the Schechter function, or $j = (6.305 \pm 0.067) \times 10^8\,$L$_\odot\,h$\,Mpc$^{-3}$ from the galaxy weights. The true luminosity density is likely to be higher than these values since the Schechter function provides a poor fit at both the faint end and the bright end, and since there are known sources of incompleteness in absolute magnitude, colour and surface brightness at the faint end. For comparison, measurements of the luminosity density from the published results shown in Fig.\ \ref{lf} lie within approximately (5.8--7.6)$\times 10^8\,$L$_\odot\,h$\,Mpc$^{-3}$ \citep[see][fig.\ 15]{Jones...2006}.

\subsection{Large-scale structure}
\label{lss}

In order to show the effect of large-scale structure on our results, we have estimated the LF for the five principal subdivisions of the sample (see Fig.\ \ref{coverage}).

\begin{figure}
\centerline {
\includegraphics[width=0.5\textwidth]{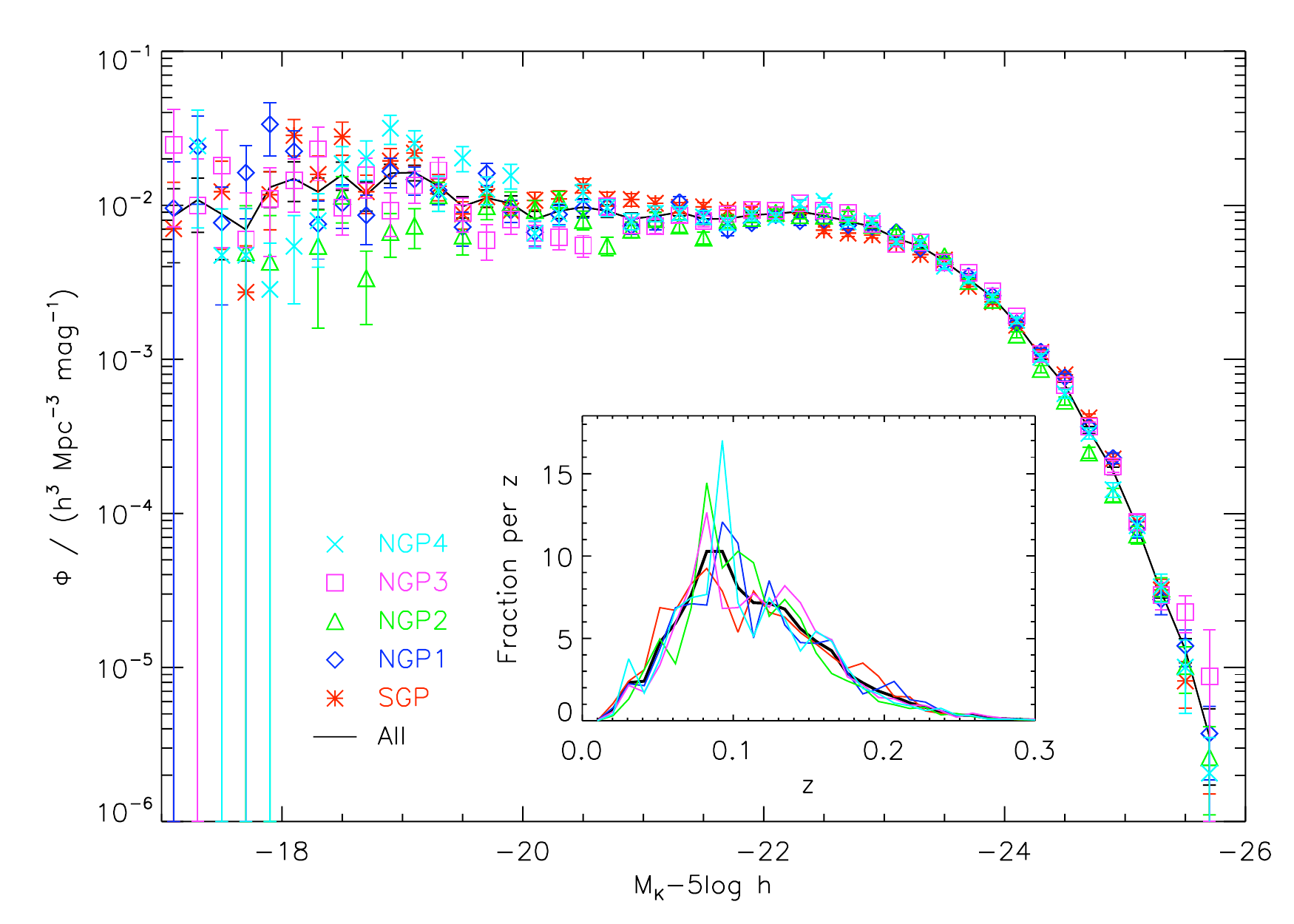}}
\caption[lf_ngpsgp]{\label{lf_ngpsgp} $K$-band LF for the NGP and SGP regions, also showing the LF for the whole sample.  Jackknife errors are shown for the whole sample and Poisson errors for the sub-samples. The inset shows the redshift distribution of galaxies in each sub-sample, with the line colours corresponding to the colours of the symbols in the main figure.}
\end{figure}

Fig.\ \ref{lf_ngpsgp} shows the $K$-band LF for the SGP region and the four NGP regions, together with the LF for the whole sample and the redshift distribution for each sub-sample. We note the following.
\begin{enumerate}
\item There is a large scatter at low luminosities, illustrating the limitations of the $1/V_\mathrm{max}$ method.
\item Each sub-sample is normalized according to the number of galaxies rather than the area covered, so a genuine over-density at a certain redshift will be artificially compensated for by an apparent under-density at other redshifts, and \textit{vice versa}.  Given the correlation between redshift and absolute magnitude (Fig.\ \ref{zhist}), this means that large-scale structure at intermediate redshifts will distort the LF at both ends. This may be seen for the SGP region, where an under-density at $z \simeq 0.1$, seen at $M_K-5\log h$ between $-22.5$ and $-23$, may have led to an over-estimate of the LF at both the faint end and the bright end, while for the NGP2 region, an over-density at $z \lesssim 0.1$ ($M_K-5\log h \simeq -22.5$), the Sloan Great Wall \citep{Gott...2005}, may have led to an under-estimate of the LF at both ends.
\item An over-density affecting the whole sample at a certain redshift will be seen at the absolute magnitude corresponding to that redshift (see Fig.\ \ref{zhist}). There is a large over-density at $z \lesssim 0.1$, which may explain why the LF in Fig.\ \ref{lf} shows a higher density at $M_K-5\log h$ between $-22$ and $-23$ compared with the findings of other authors.
\end{enumerate}

\subsection{Variation with redshift}

In order to identify further sources of bias we split the sample into three bins in redshift, containing approximately equal numbers of galaxies.

\begin{figure}
\centerline {
\includegraphics[width=0.5\textwidth]{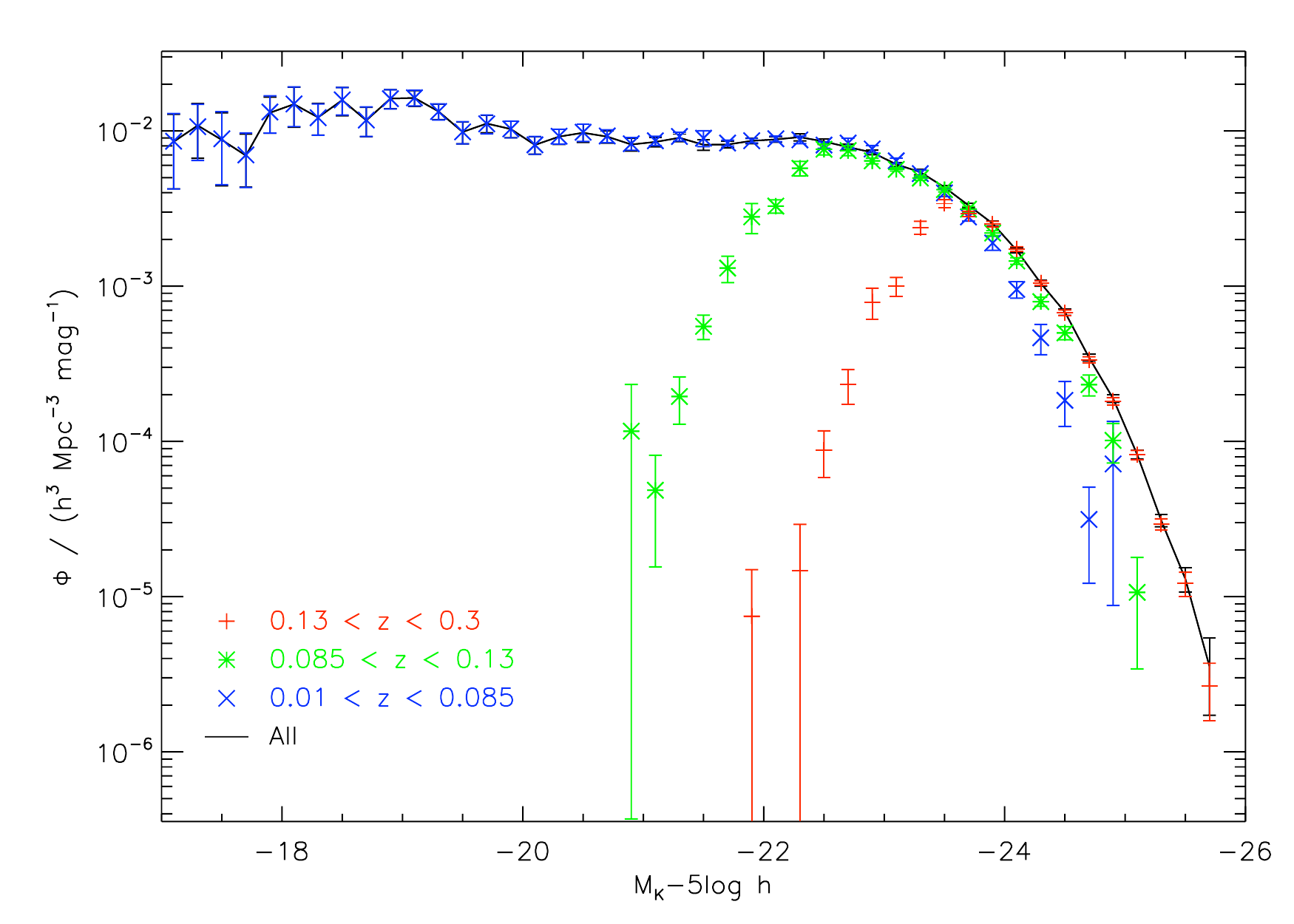}}
\caption[lf_hiloz]{\label{lf_hiloz} $K$-band LF for galaxies in three different redshift ranges, also showing the LF for the whole sample.}
\end{figure}

Fig.\ \ref{lf_hiloz} shows the $K$-band LF estimated for the low-, mid- and high-redshift sub-samples.  In the mid- and high-redshift sub-samples, there is a gradual downturn of the LF towards faint absolute magnitudes. This is to be expected, since the faint $K$-band absolute magnitude limit at any given redshift is strongly dependent on the $r-K$ colour (see Fig.\ \ref{rmk_k}). There is disagreement at the bright end between the different slices, with $\phi$ getting progressively higher at higher redshift.  This trend could be a result of the evolution corrections being too small, but the trend is still present when a correction as strong as $Q=2$ is applied.

Contributing factors could be: (1) an over-simplistic form for the evolution corrections, $E(z) = Qz$, independent of galaxy type, (2) a decreasing apparent S\'ersic index with decreasing angular size, as a result of convolution with the PSF, causing a greater fraction of a galaxy's flux to be recovered by the Petrosian magnitude when the galaxy is observed at higher redshift, (3) poorly understood limits to the sample, or (4) large-scale structure affecting both the shape and normalization of the LF for each redshift slice.

However, the most significant factor is likely to be the limit on the physical size of galaxies included in the sample, which arises through a combination of the large angular size limit and the high redshift limit (see Section \ref{sec:area}). In the medium- and low-redshift bins, the upper limit to the redshift is lowered, and hence the upper limit to the physical Petrosian radius is also lowered. From Equation \ref{eqn:radius}, the physical Petrosian radius can be no larger than $9.73h^{-1}$\,kpc in the medium-redshift bin, and no larger than $6.70h^{-1}$\,kpc in the low-redshift bin. This will cause significant incompleteness at the bright end of the LF, since higher-luminosity galaxies tend to be larger.

This test does make it clear that our results are to an extent dependent on the redshift limits chosen: choosing a lower value than 0.3 for the maximum redshift would have given an even steeper bright-end slope for the LF.

\subsection{$r$-band luminosity function and luminosity density}

\begin{figure}
\centerline {
\includegraphics[width=0.5\textwidth]{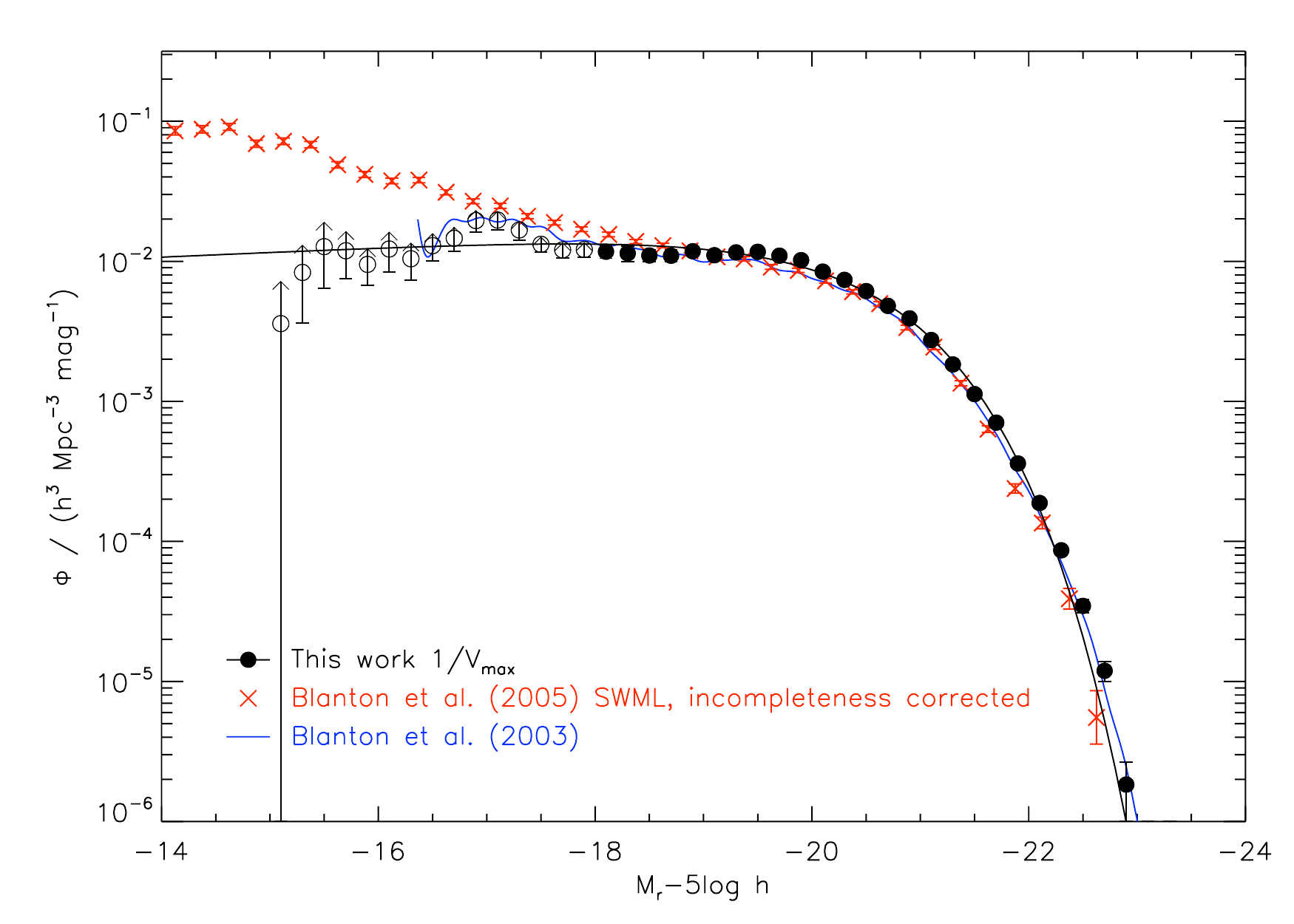}}
\caption[lf_r]{\label{lf_r} $r$-band LF\@. Only the filled points are used in the Schechter function fit, i.e., $M_r - 5 \log h$ brighter than $-18$; the unfilled points are likely to suffer from some incompleteness of low-surface brightness galaxies or low-luminosity blue galaxies. The LF of \citet{Blanton...2003c} has been adjusted from the $^{0.1}r$-band at $z=0.1$ to the $r$-band at $z=0$ by (1) shifting $Qz$\,mag fainter in magnitude, with $Q=1.62$ and $z=0.1$, (2) reducing the number density by $0.4Pz$ dex, with $P=0.18$, and (3) making the LF 0.22\,mag brighter in magnitude, to convert from $^{0.1}r$ to $r$. Schechter function parameters are $M^*-5\log h=-20.40 \pm 0.04$, $\alpha=-0.91 \pm 0.05$ and $\phi^*=(0.0195 \pm 0.0010)h^3$\,Mpc$^{-3}$. \nocite{Blanton...2005c}}
\end{figure}

Fig.\ \ref{lf_r} shows the $r$-band LF\@. While there is excellent agreement at the bright end with the LF of \citet{Blanton...2003c}, our LF is over-dense at intermediate redshifts, probably due to large-scale structure. The deficit of blue low-luminosity galaxies, identified in Fig.\ \ref{rmk_r}, is clearly evident.

The $r$-band luminosity density is found to be $j = (1.930 \pm 0.027) \times 10^8\,$L$_\odot\,h$\,Mpc$^{-3}$ by extrapolating the Schechter function, or $j = (1.931 \pm 0.019) \times 10^8\,$L$_\odot\,h$\,Mpc$^{-3}$ from the galaxy weights, assuming a solar absolute magnitude of 4.64 \citep{BlantonR2007}.  Again, the true luminosity density is likely to be higher than these values given the incompleteness at the faint end.  This is somewhat higher than the $r$-band $z=0$ luminosity density of \citet{Blanton...2003c}, $-15.90 + 2.5 \log h$\,mag in a Mpc$^3$, or $1.64 \times 10^8\,$L$_\odot\,h$\,Mpc$^{-3}$.

Given that we use the same source of data as \citet{Blanton...2003c}, but over a smaller area, with a more complex selection function and with an inferior LF estimator, the $r$-band results we find should not be interpreted as being more than a consistency check on our analysis.

\subsection{Subdividing by colour}

The bimodality of the galaxy population has been recognized by many authors \citep[see][and references therein]{Driver...2006,Ball...2006b}. This may be visualized by subdividing the LF or BBD in various ways, for example, according to colour, concentration or spectral class.  Of these properties, we find, following \citet{Driver...2006}, that the $u-r$ core (PSF) colour gives a particularly sharp dichotomy. 

\begin{figure}
\centerline {
\includegraphics[width=0.5\textwidth]{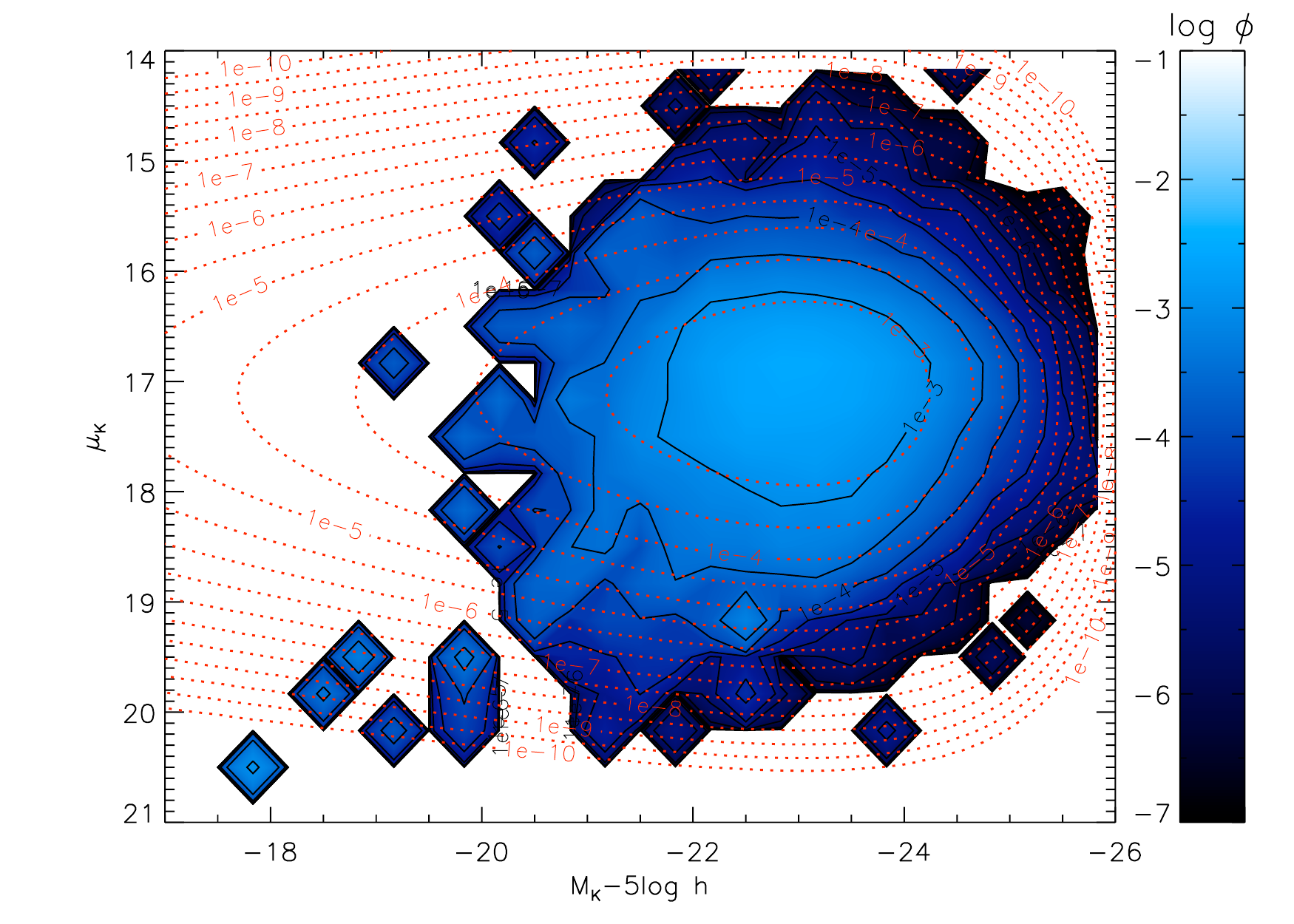}}
\caption[bbd_red]{\label{bbd_red} BBD for red galaxies, with $(u-r)_\mathrm{PSF} > 2.35$. The best-fitting Cho\l oniewski function, estimated using $M_K - 5 \log h < -20$ and $\mu_\mathrm{e,abs} < 19$, is shown by the red dotted contours. Parameters of the fit are $M^*-5\log h=-22.91$\,mag, $\alpha=0.14$,  $\phi^*=0.0119h^3$\,Mpc$^{-3}$, $\mu^*_\mathrm{e,abs}=17.11$\,mag\,arcsec$^{-2}$, $\sigma_{\mu_\mathrm{e,abs}}=0.584$\,mag\,arcsec$^{-2}$ and $\beta=-0.0006$.}
\end{figure}

\begin{figure}
\centerline {
\includegraphics[width=0.5\textwidth]{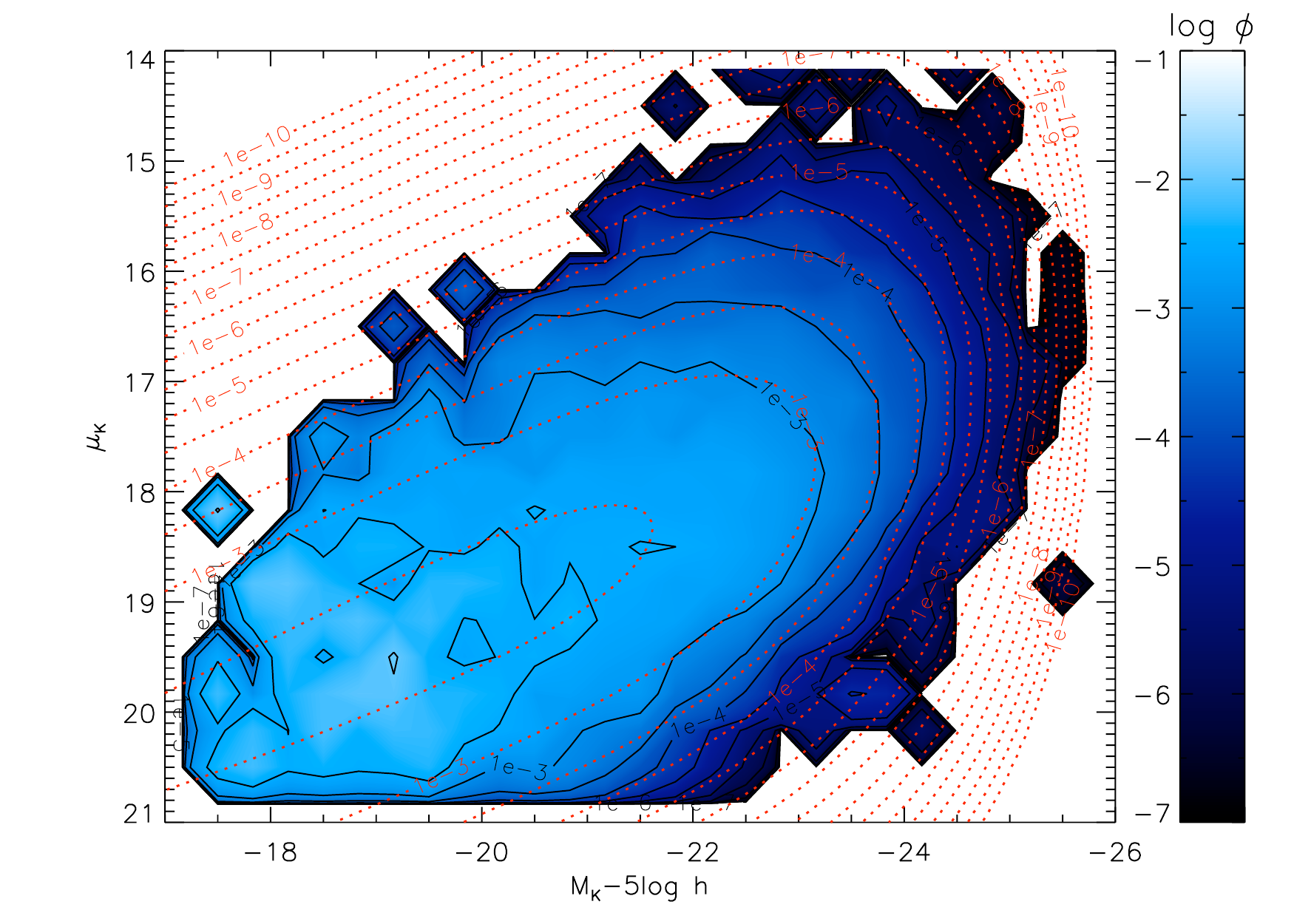}}
\caption[bbd_blue]{\label{bbd_blue} BBD for blue galaxies, with $(u-r)_\mathrm{PSF} < 2.35$. The best-fitting Cho\l oniewski function, estimated using $M_K - 5 \log h < -20$ and $\mu_\mathrm{e,abs} < 19$, is shown by the red dotted contours. Parameters of the fit are $M^*-5\log h=-22.63$\,mag, $\alpha=-0.94$,  $\phi^*=0.0115h^3$\,Mpc$^{-3}$, $\mu^*_\mathrm{e,abs}=17.91$\,mag\,arcsec$^{-2}$, $\sigma_{\mu_\mathrm{e,abs}}=0.851$\,mag\,arcsec$^{-2}$ and $\beta=0.419$.}
\end{figure}

\begin{figure}
\centerline {
\includegraphics[width=0.5\textwidth]{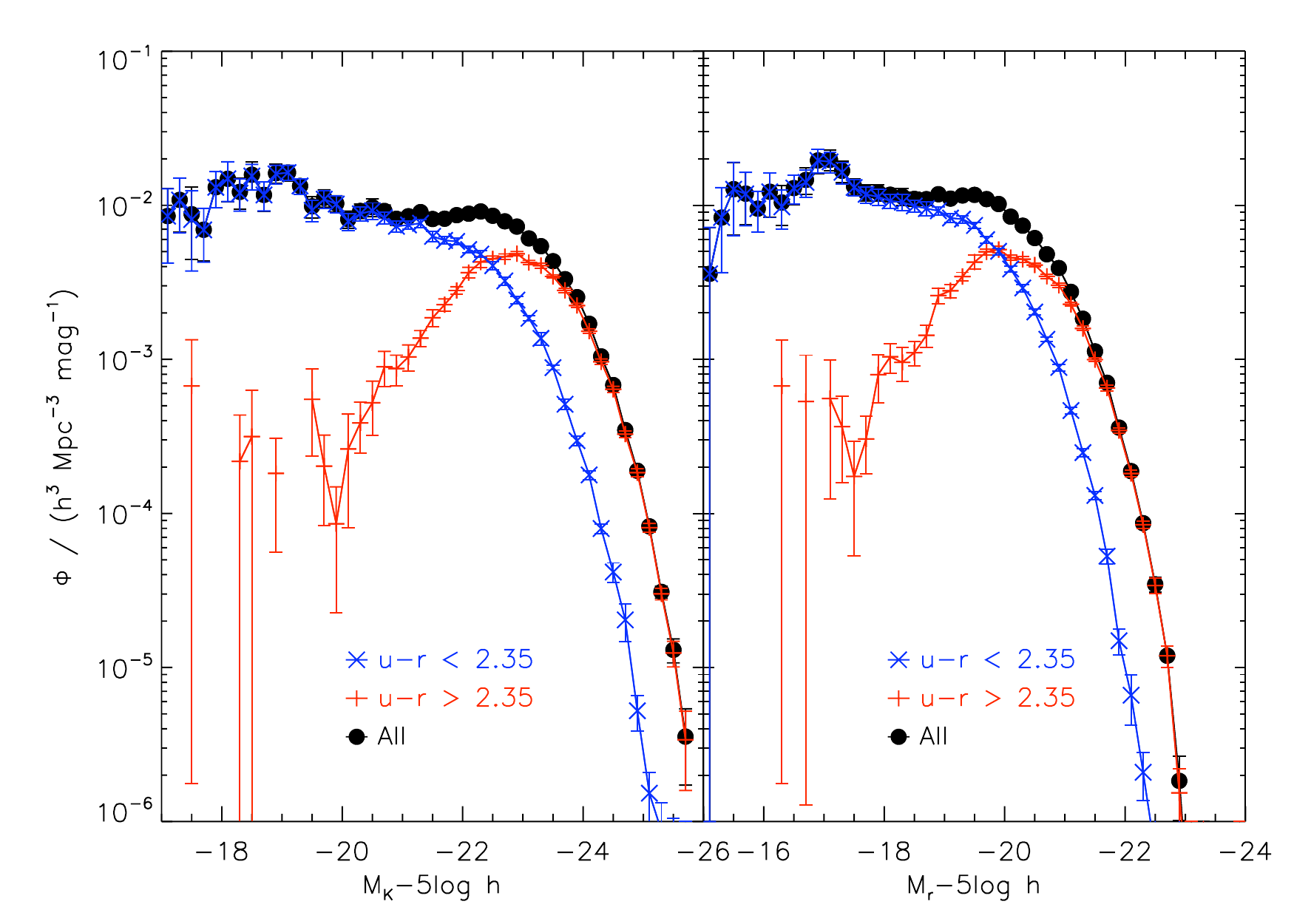}}
\caption[lf_redblue]{\label{lf_redblue} $K$- (left-hand panel) and $r$-band LFs for red and blue galaxies, showing the total LF as well.}
\end{figure}

Figs.\ \ref{bbd_red}--\ref{lf_redblue} show the $K$-band BBD and $K$- and $r$-band LFs split by the SDSS rest-frame $u-r$ PSF colour. The space density is estimated by summing the weights of galaxies with $u-r > 2.35$ or $u-r < 2.35$ for red and blue galaxies respectively, with jackknife errors estimated subsequently.

The BBD for red galaxies, excluding outliers, shows no evidence of a correlation between luminosity and surface brightness, while the BBD for blue galaxies shows no flattening off of the luminosity--surface brightness relation at high luminosities, suggesting that this division reflects a property of the underlying population. Moreover, the Cho\l oniewski function appears to fit the blue BBD much better than it fits the BBD for the whole sample. However, we caution that the lack of red-core galaxies with $M_K-5\log h > -20$ could be a symptom of the incompleteness identified in Fig.\ \ref{rmk_k}, while the lack of such galaxies with $\mu_\mathrm{e,abs} > 19.5$\,mag\,arcsec$^{-2}$ could be due to the low-surface brightness limit for de Vaucouleurs profile galaxies.

The LFs show a sharp division, with red-core galaxies more abundant than blue-core galaxies by an order of magnitude at high luminosity (and \textit{vice versa} at low luminosity), and with the bright end of the LF around 0.8\,mag more luminous in the $K$-band for red-core galaxies.

\section{Stellar mass function}

Stellar masses are derived from the \textsc{kcorrect} template fits, which are based on the $ugriz$ photometry. Described in more detail by \citet{BlantonR2007}, these templates are generated from \citet{BruzualC2003} stellar population synthesis models with a \citet{Chabrier2003b} initial mass function (IMF).

The $M/L$ ratio varies less at NIR than at optical wavelengths \citep{BelldJ2001} so, assuming the uncertainty in stellar mass is dominated by the uncertainty in the $M/L$ ratio, it makes sense to estimate the stellar mass from the NIR absolute magnitude and the $M/L$ ratio at that wavelength. Ideally the $M/L$ ratio could be found by fitting a template to all available photometry, i.e., $ugrizYJHK$ for SDSS and UKIDSS.  However, for this analysis, where we have good optical colours (with consistent apertures) but poor NIR and optical--NIR colours, we find it is best to estimate the $K$-band $M/L$ ratio using the optical colours only.

\begin{figure}
\centerline {
\includegraphics[width=0.5\textwidth]{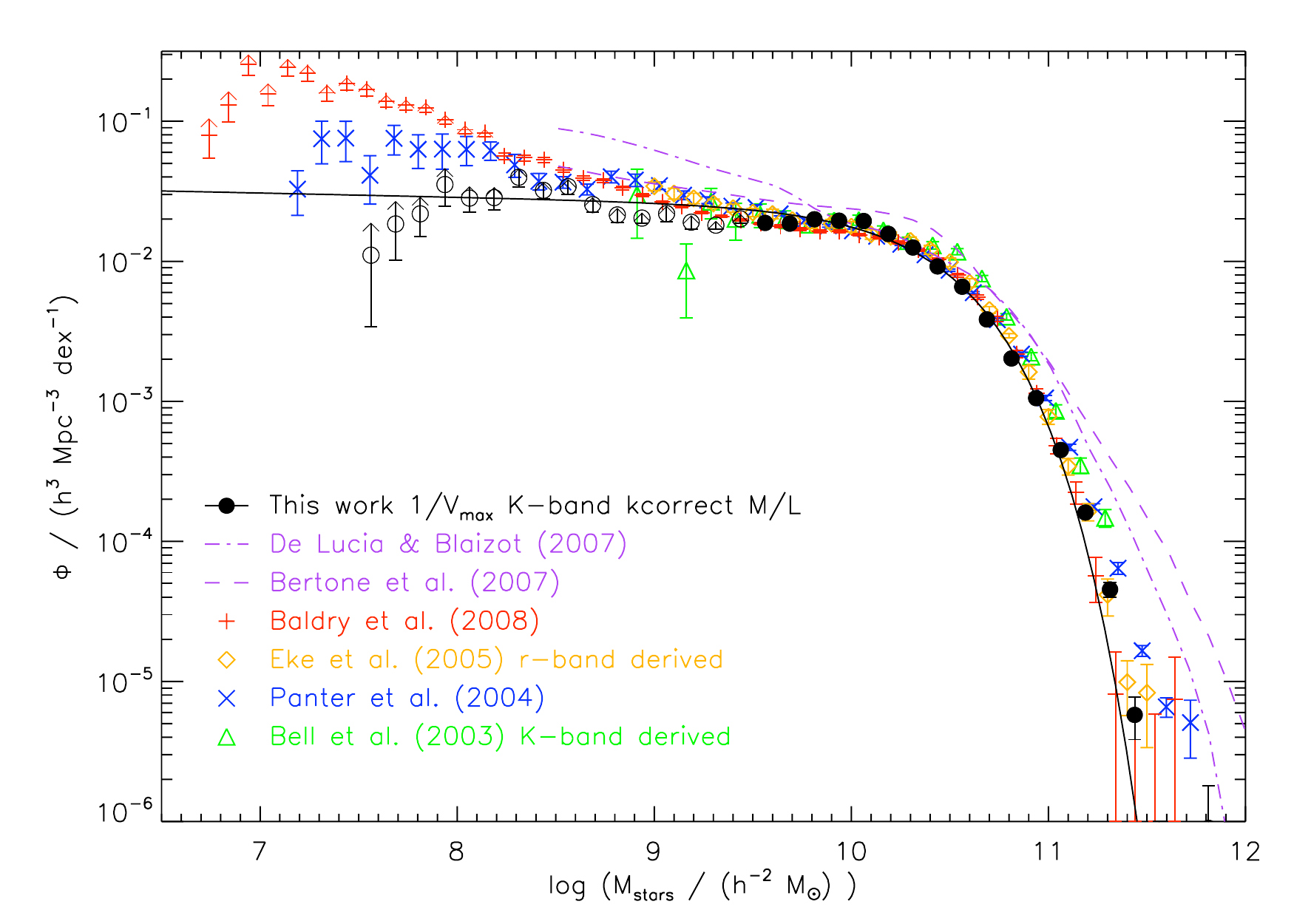}}
\caption[smf]{\label{smf} Stellar mass function. Only the filled points are used in the Schechter function fit, i.e., stellar mass greater than $10^{9.5}h^{-2} \,$M$_\odot$; the unfilled points are likely to suffer from some incompleteness of low-surface brightness galaxies. Masses calculated from the $K$-band \textsc{kcorrect} $M/L$ ratios have been increased by 0.1\,dex. Schechter function parameters are found to be $\log (\mathcal{M}^*h^2/ $M$_\odot) = 10.45 \pm 0.02$, $\alpha=-1.03 \pm 0.04$ and $\phi^*=(0.0105 \pm 0.0006)h^3$\,Mpc$^{-3}$. Stellar masses based on other IMFs have been reduced for comparison with our results, based on the Chabrier IMF: Salpeter IMF \citep*{PanterHJ2004} by 0.3\,dex, `diet' Salpeter \citep{Bell...2003c} by 0.15\,dex, and no conversion has been applied for the Kennicutt \citep{Eke...2005} or Kroupa \citep*{BaldryGD2008} IMFs. \nocite{BertonedLT2007,deLuciaB2007}}
\end{figure}

Fig.\ \ref{smf} shows the SMF, with stellar masses estimated from the $K$-band absolute magnitudes and the $K$-band $M/L$ ratios from \textsc{kcorrect}. At the high-mass end our results agree well with previously-published SMFs, while at the low-mass end the discrepancy could be a result of incompleteness or large-scale structure.  The underdensity at intermediate masses ($\mathcal{M}  \simeq 10^9 h^{-2} \,$M$_\odot$) could be due to inappropriate $M/L$ ratios or large-scale structure.

\begin{figure}
\centerline {
\includegraphics[width=0.5\textwidth]{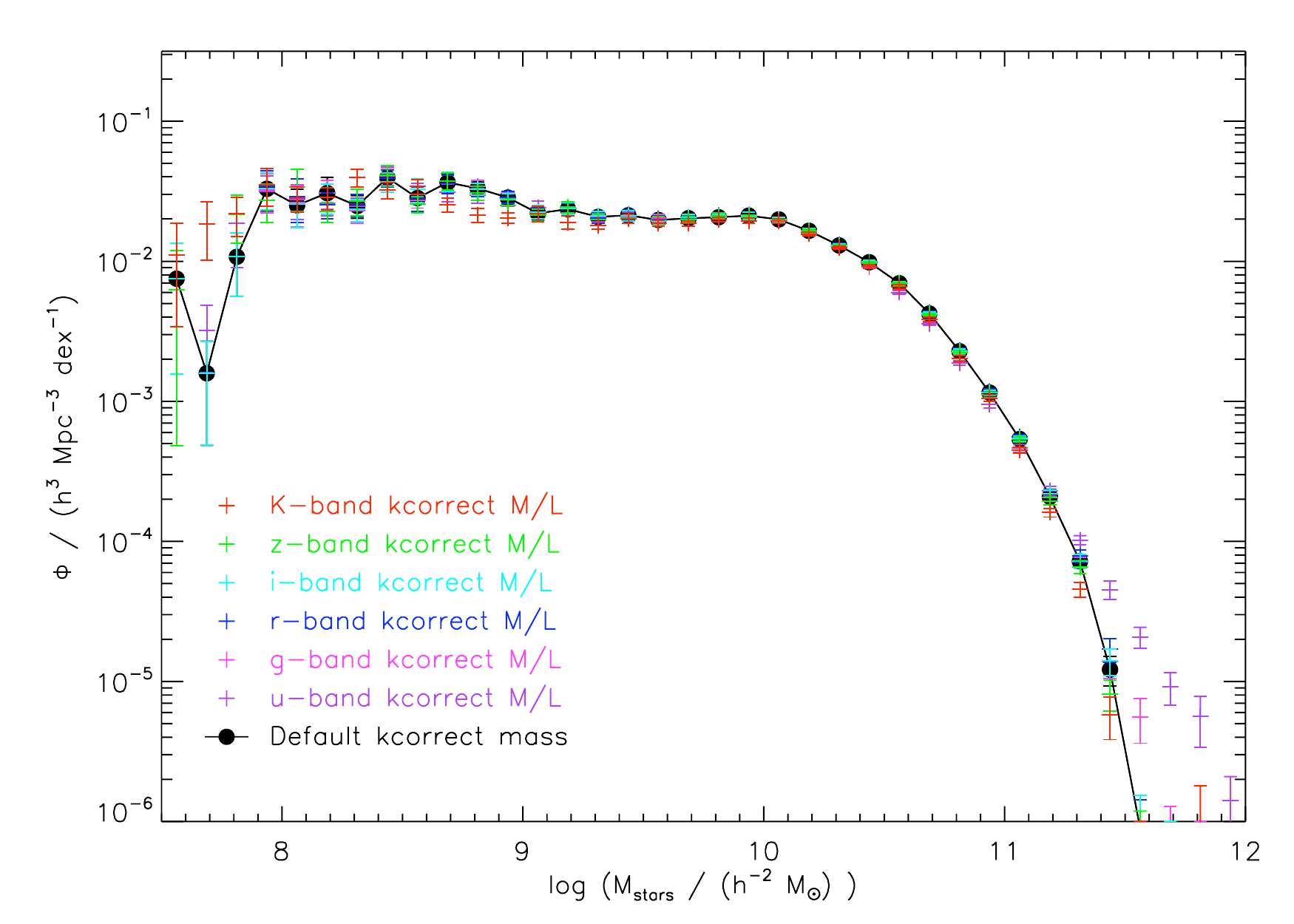}}
\caption[smf_kcorrect]{\label{smf_kcorrect} SMF, using various $M/L$ ratios and the default \textsc{kcorrect} mass, which is derived from the template fit to the input ($ugriz$) absolute magnitudes. Masses calculated from the $K$-band \textsc{kcorrect} $M/L$ ratios have been increased by 0.1\,dex.}
\end{figure}

Fig.\ \ref{smf_kcorrect} shows the SMF calculated using the default mass and the various $M/L$ ratios, all given by \textsc{kcorrect}.  We found there to be an offset in the masses derived from the $K$-band $M/L$ ratios compared with the masses derived from the optical bands. The precise cause of this offset is not known, but it has been compensated for by increasing the $K$-band-derived masses by 0.1\,dex. The greater uncertainty in blue $M/L$ ratios would be expected to stretch the high-mass end of the SMF, as seen in the $u$- and $g$-bands SMFs.  Smaller uncertainty in the red or near-infrared $M/L$ ratios may be responsible for the disagreement at low masses.  However, there is some uncertainty in the $K$-band $M/L$ ratios caused by the emission from thermally-pulsating asymptotic giant branch (TP-AGB) stars, which are very difficult to model \citep{Maraston1998,Bruzual2007}.

The stellar mass density is found to be $(3.02 \pm 0.05) \times 10^8$\,$h\,$M$_\odot$\,Mpc$^{-3}$ by extrapolating the Schechter function or $(2.99 \pm 0.04) \times 10^8$\,$h\,$M$_\odot$\,Mpc$^{-3}$ from the galaxy weights.  Due to incompleteness this is likely to be an underestimate, and a different IMF could increase this substantially, for example, by a factor of 2 for a Salpeter IMF. \citet{BaldryGD2008} find a general consensus that `the cosmic stellar mass density is in the range 4--8 per cent of the baryon density (assuming $\Omega_\mathrm{b} = 0.045$)'. In their units, taking $h=0.7$, we find a stellar mass density of $4.9 \pm 0.08$ or $4.9 \pm 0.06$ per cent of the baryon density, respectively, in good agreement with previous findings.

\section{Discussion}

Having already compared our results with those of other authors, we now consider some implications of our findings along with possible improvements and extensions.

\subsection{Functional fits}

The properties and evolution of the galaxy population are conventionally quantified using simple functional (e.g.\ Schechter function) fits to the data. With a large sample of low-redshift galaxies, it is possible to test for expected bias when such a simple form is assumed for the LF at high redshift.

\begin{figure}
\centerline {
\includegraphics[width=0.5\textwidth]{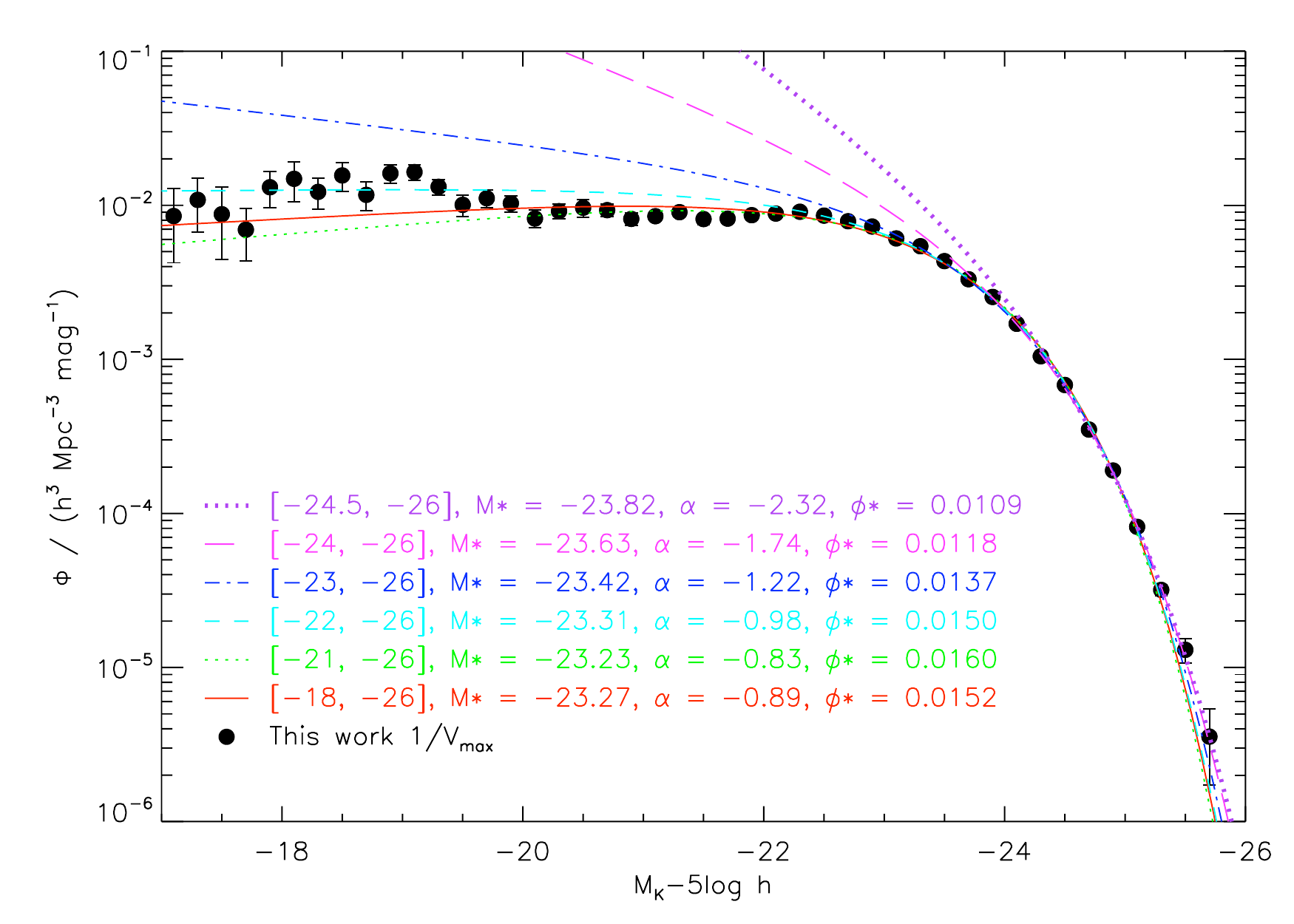}}
\caption[lf_z]{\label{lf_z} $K$-band LF, showing Schechter function fits restricting the fitting range in absolute magnitudes, intended to mimic surveys at higher redshifts.}
\end{figure}

Fig.\ \ref{lf_z} shows the Schechter function fits to our $K$-band LF, restricting the fit to various portions of the LF.  This is intended to mimic surveys at higher redshift, where only the bright end of the LF is visible.  These results suggest that, even for a non-evolving LF, with increasing redshift one would observe (1) a steeper faint-end slope, (2) a brighter characteristic magnitude and (3) a decreasing number density at the characteristic magnitude. This arises because of systematic deviations from the Schechter function form and highlights the danger of relying too strongly on the Schechter function fits.

\subsection{Possible improvements and extensions}

The results presented here use just over an eighth of the projected final overlap area between the UKIDSS LAS and SDSS (4000 deg$^2$).  Already the number of galaxies is comparable to the largest $K$-band surveys of the low-redshift universe, so we now consider how this work could be refined for a larger sample.

The $1/V_\mathrm{max}$ estimator has been shown to be sensitive to large-scale structure at low redshift.  With a larger sample this will be less apparent, but a better estimator of the space density is required, e.g., a multivariate version of the SWML method or a correction for variation with redshift \citep[e.g.,][]{Cross...2001}.  We have found the SWML method difficult to implement over four dimensions, due to the complexity of the parameter space, but these problems may not be insurmountable.

In this work the uncertainties in the measured quantities have been largely ignored and assumed to be negligible. However, the presence of magnitude errors can have a significant effect on the shape of the LF\@.  More advanced methods for estimating the LF can take this into account \citep{Blanton...2003c} but these have not yet been extended to multivariate methods.

Peculiar motions have not been considered in this work.  However, in order to probe fainter absolute magnitudes, it is necessary to reduce the low-redshift limit so that peculiar velocities can no longer be neglected.

A larger sample would allow a more comprehensive census of many galaxy properties.  It is important to have good galaxy colours, both NIR colours between the four bands in the LAS, $YJHK$, which could reflect various physical properties of the galaxies \citep{Eminian...2008}, and the optical--NIR colours, which would make template fitting possible across the nine bands in SDSS and the LAS, which in turn would lead to better $M/L$ ratios and stellar masses.  Neither of these are currently possible, since the Petrosian apertures are different in the four LAS bands, and different again to the (single) SDSS Petrosian aperture for each galaxy, and since model, e.g.\ \citet{Sersic1968} profile, magnitudes are not available. Extrapolated total magnitudes such as these would also allow a better determination of the bright end of the LFs and SMF.

\citet{Sersic1968} profile fits would also provide structural information about the galaxies, as would bulge-to-disc decomposition \citep{Driver...2007a,Driver...2007b}.

The limits in radius and surface-brightness also require attention.  A deficit in low-surface brightness galaxies can be accounted for by assuming a certain distribution, e.g.\ Gaussian in surface brightness in bins of absolute magnitude, and then extrapolating.  But it is likely that a lower surface brightness limit could be reached using source extraction optimized for low-redshift extended sources, for example, using elliptical apertures. Forthcoming deep near-infrared surveys, such as the VISTA (Visible and Infrared Survey Telescope for Astronomy) Kilo-degree Infrared Galaxy (VIKING) survey, are anticipated to provide measurements to a fainter surface brightness limit.

The colour-dependent incompleteness at low luminosity is a result of having a survey limited in both $r$ and $K$.  This could be improved by supplementing SDSS redshifts with spectroscopy of galaxies within some $K$-band completeness limit.  Redshifts from the recently started GAMA\footnote{Galaxy And Mass Assembly, \url{http://www.eso.org/~jliske/gama/}} spectroscopic survey could be used to probe regions of the parameter space not currently sampled.

\section{Conclusions}

We have presented the first statistical analysis of galaxies from the UKIDSS LAS\@.  The $1/V_\mathrm{max}$ space density estimator has been used in a four-dimensional form to produce results for the $K$- and $r$-band LF and the SMF consistent with previous findings, with a form similar to a Schechter function with an almost flat faint-end slope.

We have presented the first $K$-band BBD in $K$-band absolute magnitude and effective surface brightness. This shows similar trends to the optical BBD: a correlation between luminosity and surface brightness, with a broadening of the surface brightness distribution at low luminosity and a flattening of the luminosity--surface brightness relation at high luminosity.

The multiple limits on the survey have been taken into account. For example, limits in $K$-band and $r$-band magnitude, $K$-band Petrosian radius and $K$-band surface brightness have been used to estimate the volume within which each galaxy would have been visible.

When the sample is subdivided according to colour, we have found a clear distinction between two populations, one of red, high-luminosity and high-surface-brightness galaxies, and the other of blue, low-luminosity and low-surface-brightness galaxies. This agrees well with previous investigations of the bimodality of the galaxy population.

We have considered various sources of incompleteness of the data and find that in order to obtain better results for low luminosity galaxies with UKIDSS, various selection effects need further attention.  A stable multivariate estimator for the space density needs to be developed that takes consideration of large-scale structure.  Incompleteness at low surface brightness also needs to be addressed, perhaps using source extraction with elliptical apertures, and the colour-dependent selection effects could be addressed by a $K$-band limited redshift survey to extend the SDSS spectroscopic sample to a greater depth in $K$. Physical properties of galaxies could be investigated further with good NIR and optical--NIR colours and/or extrapolated model magnitudes.

\section*{Acknowledgments}

We thank Seb Oliver for useful discussions, Simon Driver and C\'eline Eminian for comments on an earlier draft of this paper, Nigel Hambly and others at the Edinburgh WFAU for assistance with the UKIDSS data and Serena Bertone for providing model luminosity and stellar mass functions.  We are grateful to Nigel Metcalfe for providing number count data from the literature on his web page. We thank the anonymous referee for helpful and constructive comments. AJS has been supported by an STFC grant.

This work is based largely on data obtained as part of the UKIRT Infrared Deep Sky Survey.

Funding for the SDSS and SDSS-II has been provided by the Alfred P.\ Sloan Foundation, the Participating Institutions, the National Science Foundation, the U.S. Department of Energy, the National Aeronautics and Space Administration, the Japanese Monbukagakusho, the Max Planck Society, and the Higher Education Funding Council for England. The SDSS Web Site is \url{http://www.sdss.org/}.

The SDSS is managed by the Astrophysical Research Consortium for the Participating Institutions.  The Participating Institutions are the American Museum of Natural History, Astrophysical Institute Potsdam, University of Basel, Cambridge University, Case Western Reserve University, University of Chicago, Drexel University, Fermilab, the Institute for Advanced Study, the Japan Participation Group, Johns Hopkins University, the Joint Institute for Nuclear Astrophysics, the Kavli Institute for Particle Astrophysics and Cosmology, the Korean Scientist Group, the Chinese Academy of Sciences (LAMOST), Los Alamos National Laboratory, the Max-Planck-Institute for Astronomy (MPIA), the Max-Planck-Institute for Astrophysics (MPA), New Mexico State University, Ohio State University, University of Pittsburgh, University of Portsmouth, Princeton University, the United States Naval Observatory, and the University of Washington.

This research has made use of NASA's Astrophysics Data System.


\label{lastpage}

\end{document}